\documentclass[a4paper,11pt]{article}
\pdfoutput=1
\usepackage{jheppub}
\usepackage[T1]{fontenc}
\usepackage{multirow, makecell}
\usepackage{graphics}
\usepackage{pifont}
\usepackage{dcolumn}
\newcolumntype{d}[1]{D{.}{.}{#1}}
\usepackage{booktabs}
\usepackage{xspace}

\newcommand{\GeV}{\mathrm{GeV}}
\DeclareRobustCommand{\cjet}{\ensuremath{c}\text{-jet}\xspace}
\DeclareRobustCommand{\cjets}{\ensuremath{c}\text{-jets}\xspace}
\DeclareRobustCommand{\WPC}{\ensuremath{W^+}+\cjet}
\DeclareRobustCommand{\WMC}{\ensuremath{W^-}+\cjet}
\DeclareRobustCommand{\WC}{\ensuremath{W^\pm}+\cjet\xspace}
\DeclareRobustCommand{\RAT}{\ensuremath{\mathcal{R}^\pm_c}\xspace}

\DeclareRobustCommand{\yl}{\ensuremath{|y_\ell|}\xspace}
\DeclareRobustCommand{\etac}{\ensuremath{|\eta_{j_c}|}\xspace}
\DeclareRobustCommand{\ptc}{\ensuremath{p_{T,j_c}}\xspace}
\DeclareRobustCommand{\etmiss}{\ensuremath{E_{T,{\rm miss}}}\xspace}
\DeclareRobustCommand{\ptl}{\ensuremath{p_{T,\ell}}\xspace}
\DeclareRobustCommand{\etw}{\ensuremath{E_{T,W}}\xspace}

\DeclareRobustCommand{\sb}{\ensuremath{\bar{s}}\xspace}
\DeclareRobustCommand{\cb}{\ensuremath{\bar{c}}\xspace}
\DeclareRobustCommand{\qb}{\ensuremath{\bar{q}}\xspace}

\preprint{{\raggedleft%
ZU-TH 76/23 \\
IPPP/23/69 \\
CERN-TH-2023-217 \\
}}

\title{Precise QCD predictions for W-boson production in association with a charm jet}

\author[a,b]{A.~Gehrmann--De Ridder,}
\author[b]{T.~Gehrmann,}
\author[c,d]{E.~W.~N.~Glover,}
\author[e]{A.~Huss,}
\author[a]{A.~Rodriguez Garcia,}
\author[b,f]{G.~Stagnitto}

\affiliation[a]{Institute for Theoretical Physics, ETH, CH-8093 Z\"urich, Switzerland}
\affiliation[b]{Department of Physics, University of Z\"urich, CH-8057 Z\"urich, Switzerland}
\affiliation[c]{Institute for Particle Physics Phenomenology, Durham University,  Durham DH1 3LE, UK}
\affiliation[d]{Department of Physics, Durham University,  Durham DH1 3LE, UK}
\affiliation[e]{Theoretical Physics Department, CERN, CH-1211 Geneva 23, Switzerland}
\affiliation[f]{Universit\`{a} degli Studi di Milano-Bicocca \& INFN, Piazza della Scienza 3, I-20126 Milano, Italy}

\emailAdd{gehra@phys.ethz.ch}
\emailAdd{thomas.gehrmann@uzh.ch}
\emailAdd{e.w.n.glover@durham.ac.uk}
\emailAdd{alexander.huss@cern.ch}
\emailAdd{adrianro@phys.ethz.ch}
\emailAdd{giovanni.stagnitto@unimib.it}

\abstract{%
The production of a $W$-boson with a charm quark jet provides a highly
sensitive probe of the strange quark distribution in the proton. Employing a
novel flavour dressing procedure to define charm quark jets, we compute
$W$+charm-jet production up to next-to-next-to-leading order (NNLO) in QCD.
We study the perturbative stability of production cross sections with
same-sign and opposite-sign charge combinations for the $W$ boson and the
charm jet. A detailed breakdown according to different partonic initial states
allows us to identify particularly suitable observables for the study of the
quark parton distributions of different flavours.%
}

\begin{document} 
\maketitle
\flushbottom

\section{Introduction}
\label{sec:intro}

The quark and gluon content of the proton is described by parton distributions
functions (PDFs), which parametrise the probabilities for a given parton species
to carry a specific fraction of the longitudinal momentum of a fastly moving
proton. PDFs can not be computed from first principles in perturbative QCD,
which determines only their evolution with the resolution
scale~\cite{Altarelli:1977zs,Dokshitzer:1977sg}. The initial distributions for
all quark and antiquark flavours and gluons are thus determined from global
fits~\cite{Gao:2017yyd,Alekhin:2017kpj,Hou:2019efy,Bailey:2020ooq,NNPDF:2021njg}
to a large variety of experimental data from high-energy collider and
fixed-target experiments. The resulting PDFs do not have uniform uncertainties
across the different quark flavours, since only some flavour combinations are
tightly constrained by precision data, e.g.\ from inclusive neutral-current
structure functions or from vector boson production cross sections. In
particular the strange quark and antiquark distributions are mainly constrained
from fixed-target neutrino-nucleon scattering
data~\cite{Botts:1993shg,NuTeV:2005wsg}.

The production of a massive gauge boson in association with a flavour-identified
jet offers a unique possibility to study PDFs for specific quark flavours.
$W$+charm-jet
production~\cite{Baur:1993zd,Giele:1995kr,Stirling:2012vh,Czakon:2020coa,Czakon:2022khx}
is of particular relevance, since its Born-level production cross section is
largely dominated by initial states colliding a gluon and a strange quark. By
selecting the $W$ charge, strange and anti-strange distributions can be probed
separately.  The production of $W$ bosons with heavy quarks has been studied by
ATLAS~\cite{ATLAS:2023ibp}, CMS~\cite{CMS:2013wql,CMS:2021oxn,CMS:2023aim} and
LHCb~\cite{LHCb:2015bwt}. However, these measurements use various different
prescriptions to identify the presence of the heavy flavour, such as for example
by tagging a specific heavy hadron species, or by a flavour-tracking in the jet
clustering.

The definition and identification of jet flavour~\cite{Banfi:2006hf} is highly
non-trivial due to possible issues with infrared and collinear safety (IRC)
related to the production of secondary quark-antiquark pairs that can partially
or fully contribute to the jet flavour. Several proposals to assign flavour to
jets in an IRC safe were recently put
forward~\cite{Caletti:2022hnc,Czakon:2022wam,Gauld:2022lem,Caola:2023wpj}, and a
generic prescription to test the IRC safety of jet flavour definitions has been
formulated~\cite{Caola:2023wpj}.
 
To include precision data from $W$+charm production processes in global PDF
fits, higher-order QCD corrections to the respective production cross sections
are required. These have been computed previously for \mbox{$W$+charm-jet}
production to next-to-next-to-leading order
(NNLO)~\cite{Czakon:2020coa,Czakon:2022khx}, while $W$+charm-hadron production
is currently only known to next-to-leading order (NLO) by combining the
identified quark production at this order with a parton-shower and
hadronization model~\cite{Bevilacqua:2021ovq,FerrarioRavasio:2023kjq}.

In this paper, we present a new NNLO computation of $W$+charm-jet production,
employing the flavour dressing procedure~\cite{Gauld:2022lem} to define charm
quark jets. Our calculation is performed in the NNLOJET parton-level event
generator framework~\cite{Gehrmann-DeRidder:2015wbt}, which implements the
antenna subtraction
method~\cite{Gehrmann-DeRidder:2005btv,Daleo:2006xa,Currie:2013vh} for the
handling of infrared singular real radiation configurations up to NNLO.  Using
this new implementation, we investigate the effects of higher-order QCD
corrections on different charge-identified $W$+charm-jet cross sections and
kinematical distributions. We decompose the predictions according to the
partonic composition of the initial state, which allows us to quantify the
sensitivity of different types of observables on the PDFs of strange quarks and
of other quark flavours.

The paper is structured as follows. In Section~\ref{sec:calculation}, we
describe the calculation of the NNLO QCD corrections, elaborating in particular
on the extensions to antenna subtraction and to the NNLOJET code required for
flavour and charge tracking. Section~\ref{sec:results} describes the results for
the flavour and charge identified distributions at NNLO and investigates their
perturbative stability. We perform a detailed decomposition into partonic
channels in Section~\ref{sec:chbreak} and discuss various observations that can
be made based on this channel breakdown. We conclude with a summary in
Section~\ref{sec:concl}.

\section{Details of the calculation}
\label{sec:calculation}

Our calculation of the NNLO corrections to $W+c$-jet production is based on the
NNLOJET parton-level event generator framework, which implements the antenna
subtraction method~\cite{Gehrmann-DeRidder:2005btv,Daleo:2006xa,Currie:2013vh}
for the cancellation of infrared singular terms between real radiation and
virtual contributions. It builds upon the NNLOJET implementation of $W+$jet
production~\cite{Gehrmann-DeRidder:2017mvr,Gehrmann-DeRidder:2019avi}.  The NNLO
corrections consist of three types of contributions: two-loop virtual (double
virtual, VV), single real radiation at one loop (real-virtual, RV) and double
real radiation (RR). The matrix elements for these contributions to $W+$jet
production are well-known and can be expressed in compact analytic
form~\cite{Garland:2001tf,Garland:2002ak,Glover:1996eh,Campbell:1997tv,Bern:1997sc,Hagiwara:1988pp,Berends:1988yn}.

The $W+$jet implementation in NNLOJET had to be extended in various aspects to
enable predictions for jets containing an identified charm quark, as described
in detail in the following subsections. The full dependence of the subprocess
matrix elements on the initial- and final-state quark flavours (including CKM
mixing effects) had to be specified, a flavour dressing procedure for the
assignment of jet flavour~\cite{Gauld:2022lem} and the flavour tracking in all
stages of the calculation had to be implemented, and the antenna subtraction
terms had to be adapted to allow for full flavour and charge tracking.

\subsection{Implementation of CKM flavour mixing}

Quark flavour mixing effects in processes involving final-state $W^\pm$ bosons
were previously included in NNLOJET by constructing CKM-weighted combinations of
incoming parton luminosities. This prescription allowed to minimise the number
of evaluations of subprocess matrix elements and associated subtraction terms
per phase space point, thereby contributing to the numerical efficiency of the
calculation. This implementation relies on a flavour-agnostic summation over all
final-state quarks and antiquarks, and does not allow to assign a specific quark
flavour to any final state object.

In the case of $Z+b$ production~\cite{Gauld:2020deh} and $Z+c$
production~\cite{Gauld:2023zlv}, the respective final-state quark flavours could
be extracted, starting from the $Z$+jet matrix elements, in a rather
straightforward manner by excluding them from the flavour sum, and keeping the
identified flavour contribution as a separate process. For $W+c$ production,
flavour identification required to dress all matrix elements with the respective
CKM factors at the $W$ interaction vertex, thereby fixing the associated quark
flavours in the initial and final state.  Where appropriate, initial state
flavour combinations were again concatenated into weighted combinations of
parton luminosities for computational efficiency, while final-state flavours
(and quark charges) were clearly identified for all subprocesses.

\subsection{Flavour dressing of jets and flavour tracking in NNLOJET}

In order to compute observables sensitive to the flavour of the particles
involved, it is necessary to retain the flavour information in both matrix
elements and subtraction terms.
A mechanism of flavour tracking has been implemented in NNLOJET,
see~\cite{Gauld:2019yng} for an overview of this procedure.
Here we stress the fact that the reduced matrix elements within the same
subtraction term can have different flavour structures, because they are related
to different unresolved limits of the matrix element.
This observation will be crucial in Section~\ref{sec:chtrack} below.

Once we have the flavour information of final-state particles at our disposal,
it is important to adopt an infrared and collinear (IRC) safe definition of
flavour of hadronic jets.
In other words, we require that the flavour of jets is not affected by the
emission of soft particles and/or collinear splittings (e.g.\ $g \to c\bar{c}$),
in order to guarantee the local cancellation of singularities between matrix
elements and subtraction terms.
Several proposals to assign flavour to jets in an IRC safe way have recently
appeared~\cite{Caletti:2022hnc,Czakon:2022wam,Gauld:2022lem,Caola:2023wpj}.
In the present analysis, we will adopt the flavour dressing algorithm
of~\cite{Gauld:2022lem}.
The key property of this approach is that the flavour assignment of jets is
entirely factorised from the initial jet reconstruction.
Hence, we can define the flavour of anti-$k_t$ jets---the de facto standard at
the LHC---in an IRC safe way.

However, in Ref.~\cite{Caola:2023wpj} it has been shown that the original
formulation of the flavour dressing algorithm as presented
in~\cite{Gauld:2022lem} starts being IRC unsafe at higher orders.
This has been proven by looking at explicit partonic configurations with many
hard and soft/collinear particles and by developing a dedicated numerical
framework for fixed-order tests of IRC safety.

After the findings of~\cite{Caola:2023wpj}, the flavour dressing algorithm has
been adjusted, and the new version passes the numerical fixed-order tests
of~\cite{Caola:2023wpj} up to $\mathcal{O}(\alpha_s^6)$.  In the new
formulation, flavoured {\em clusters} are no longer used; instead, all particles
directly enter the flavour assignment step, and we run a sequential
recombination algorithm by considering both distances between particles and
between particles and jets.

\subsection{Charge tracking in quark-antiquark antenna functions}\label{sec:chtrack}

Previous NNLOJET calculations of $Z+b$ production~\cite{Gauld:2020deh} and $Z+c$
production~\cite{Gauld:2023zlv} always summed over the charges of the identified
quarks, i.e.\ $q=(b,c)$ could be either a flavour-identified quark or a
flavour-identified antiquark. Furthermore, in any given subprocess, quarks and
antiquarks of the same flavour always come in pairs in these calculations.  In
the current calculation of $W+c$-jet production in NNLOJET, this is no longer
the case, since a charm quark that has a direct coupling to the $W$-boson will
be associated with its corresponding isospin partner (predominantly the strange
antiquark $\bar{s}$, or the CKM-suppressed down-antiquark $\bar{d}$). Moreover,
it is desirable to be able to distinguish charm quarks and antiquarks, thereby
allowing the study of charge correlations between the produced $W$ boson and the
identified charm (anti-)quark (same-sign, SS, and opposite-sign, OS,
observables), as is done in the experimental analyses.

This charge identification requires a slight extension of the antenna
subtraction formalism to accommodate the charge-tracking in the quark-antiquark
antenna functions. The requirement of charge-tracking can be illustrated with an
example. We consider the gluon-induced double real radiation contribution to
$W^- c$ production: $$g(p_1)g(p_2)\to W^-(q) c(p_i)\bar s(p_j) g(p_k),$$ which
contains the colour-ordered subprocess matrix element:
\begin{equation}
\tilde{B}_{3,W^-}^0(i_c, 1_g,k_g,2_{\tilde{g}} ,j_{\bar{s}}),
\label{eq:B30W}
\end{equation}
at first subleading colour level.  Here $\tilde{g}$ denotes the abelian-like
gluon that is colour-connected only to the quark-antiquark pair, while the other
two gluons are colour-connected to each other and to either the quark or the
antiquark.  The partonic labelling of the momenta is in all-final kinematics,
with incoming particles denoted by momenta 1 and 2.

The subtraction of triple-collinear limits corresponding to the splitting of the
incoming (non-abelian) gluon into a quark-antiquark-gluon cluster (from which
either the quark or the antiquark enters the hard subprocess) requires the
leading-colour quark-antiquark antenna function
$A_4^0(i_q,1_g,k_g,j_{\bar{q}})$.  This antenna function contains two triple
collinear limits: TC($q_i\parallel g_1 \parallel g_k$) and
TC($\bar{q}_j\parallel g_k \parallel g_1$). The associated triple-collinear
splitting functions correspond to different colour orderings and are not
identical. In these two limits, \eqref{eq:B30W} factorises as follows:
\begin{eqnarray}
\tilde{B}_{3,W^-}^0(i_c, 1_g,k_g,2_{\tilde{g}},j_{\bar{s}}) &\stackrel{i\parallel 1 \parallel k}{\longrightarrow}& P_{q_i\parallel g_1 \parallel g_k} B_{1,W^-}^0(\hat{1}_c,2_g,j_{\bar{s}}) \, ,
\nonumber \\
\tilde{B}_{3,W^-}^0(i_c, 1_g,k_g,2_{\tilde{g}},j_{\bar{s}}) &\stackrel{j\parallel k \parallel 1}{\longrightarrow}& P_{ \bar{q}_j\parallel g_k \parallel g_1} B_{1,W^-}^0(i_c,2_g,\hat{1}_{\bar{s}}) \, ,
\end{eqnarray}
where $\hat{1}$ denotes the composite momentum that flows into the hard matrix
element after the collinear splitting.  It becomes evident that only the
$\bar{q}_j\parallel g_k \parallel g_1$ limit factorises onto a matrix element
corresponding to a $W^-c$ final state, while the $q_i\parallel g_1 \parallel
g_k$ leads to a $W^- \bar{s}$ final state with an anti-charm quark in the
initial state of the reduced matrix element.  To construct the RR subtraction
term for~\eqref{eq:B30W}, one must therefore split
$A_4^0(i_q,1_g,k_g,j_{\bar{q}})$ into sub-antenna functions that contain only a
well-defined subset of its infrared limits.  The split is analogous to the split
that is used for the initial-final quark-antiquark antenna function at
NLO~\cite{Gehrmann-DeRidder:2005btv,Daleo:2006xa}:
\begin{equation}
A_3^0(i_q,1_g,j_{\bar{q}}) = a_3^0(i_q,1_g,j_{\bar{q}}) + a_3^0(j_{\bar{q}},1_g,i_q), 
\end{equation}
where 
$a_3^0(i_q,1_g,j_{\bar{q}})$ contains only the $q_i\parallel g_1$ collinear limit. 

The decomposition into sub-antennae reads as follows:
\begin{equation}
A_4^0(i_q,1_g,k_g,j_{\bar{q}})  = a_4^{0,c}(i_q,1_g,k_g,j_{\bar{q}}) + a_4^{0,d}(i_q,1_g,k_g,j_{\bar{q}}),  
\label{eq:A40dec}
\end{equation}
where we require $a_4^{0,c}$ to contain all limits where the incoming gluon
$1_g$ becomes collinear to quark $i_q$ and $a_4^{0,d}$ to contain all limits
where it becomes collinear to antiquark $j_{\bar{q}}$. Consequently, these
sub-antenna functions should contain the following double unresolved (triple
collinear, TC, double single collinear, DC, and soft-collinear, SC) limits:
\begin{eqnarray} 
a_4^{0,c}(i_q,1_g,k_g,j_{\bar{q}})  &\supset& \mathrm{TC}(q_i\parallel g_1 \parallel g_k), 
\mathrm{DC}(q_i\parallel g_1, \bar{q}_j\parallel g_k), \mathrm{SC}(q_i\parallel g_1,g_k\mbox{ soft})\,, \nonumber \\
a_4^{0,d}(i_q,1_g,k_g,j_{\bar{q}})  &\supset & \mathrm{TC}(\bar{q}_j\parallel g_k \parallel g_1) \,. \nonumber 
\end{eqnarray}
The behaviour in the single unresolved limits is more complicated, since the
sub-antenna functions should factor onto appropriate three-parton antenna
functions $A_3^0$ or their respective sub-antennae:
\begin{eqnarray}
a_4^{0,c}(i_q,1_g,k_g,j_{\bar{q}})  &\stackrel{i\parallel 1}{\longrightarrow}& P_{q_i\parallel g_1} A_3^0(\hat{1}_q,k_g,j_{\bar{q}})\,, \nonumber \\
a_4^{0,d}(i_q,1_g,k_g,j_{\bar{q}})  &\stackrel{i\parallel 1}{\longrightarrow}& 0 \,,\nonumber \\
a_4^{0,c}(i_q,1_g,k_g,j_{\bar{q}})  &\stackrel{k\parallel 1}{\longrightarrow}& P_{g_k\parallel g_1} a_3^0(i_q,\hat{1}_g,j_{\bar{q}}) \,, \nonumber \\
a_4^{0,d}(i_q,1_g,k_g,j_{\bar{q}})  &\stackrel{k\parallel 1}{\longrightarrow}& P_{g_k\parallel g_1} a_3^0(j_{\bar{q}},\hat{1}_g,i_q) \,, \nonumber \\
a_4^{0,c}(i_q,1_g,k_g,j_{\bar{q}})  &\stackrel{k\parallel j}{\longrightarrow}& P_{q_j\parallel g_k} a_3^0(i_q,1_g,(jk)_{\bar{q}}) \,, \nonumber \\
a_4^{0,d}(i_q,1_g,k_g,j_{\bar{q}})  &\stackrel{k\parallel j}{\longrightarrow}& P_{q_j\parallel g_k} a_3^0(j_{\bar{q}},1_g,(jk)_q) \,, \nonumber \\
a_4^{0,c}(i_q,1_g,k_g,j_{\bar{q}})  &\stackrel{k\mbox{ soft}}{\longrightarrow}& S_{1kj} a_3^0(i_q,1_g,i_{\bar{q}})\,,  \nonumber \\
a_4^{0,d}(i_q,1_g,k_g,j_{\bar{q}})  &\stackrel{k\mbox{ soft}}{\longrightarrow}& S_{1kj} a_3^0(j_{\bar{q}},1_g,i_q)\,,  \label{eq:a40singlelim}
\end{eqnarray}
where $(jk)$ denotes the momentum of the collinear final-state cluster, $P$ are
the collinear splitting factors and $S$ are eikonal factors.

The decomposition~\eqref{eq:A40dec} of $A_4^0(i_q,1_g,k_g,j_{\bar{q}})$ into its
sub-antennae starts from its triple collinear behaviour. The triple collinear
limit TC($q_i\parallel g_1 \parallel g_k$) is characterised by the Mandelstam
invariants $(s_{i1k},s_{i1},s_{1k},s_{ik})$ becoming simultaneously small, while
the TC($\bar{q}_j\parallel g_k \parallel g_1$) corresponds to
$(s_{1kj},s_{1k},s_{kj},s_{1j})$ becoming small. From these sets, $s_{ik}$ and
$s_{1j}$ do not appear as denominators in $ A_4^0(i_q,1_g,k_g,j_{\bar{q}})$ due
to its colour-ordering. Any denominator containing $s_{i1k}$ or $s_{i1}$ is then
partial fractioned against any denominator with $s_{1kj}$ or $s_{1k}$, using
e.g.
\begin{equation}
\frac{1}{s_{i1k}s_{1kj}} = \frac{1}{s_{i1k}(s_{i1k}+s_{1kj})}+ \frac{1}{s_{1kj}(s_{i1k}+s_{1kj})}\, ,
\end{equation}
followed by a power-counting to assign terms that are sufficiently singular (two
small invariants) in TC($q_i\parallel g_1 \parallel g_k$) to
$a_4^{0,c}(i_q,1_g,k_g,j_{\bar{q}})$ and terms from TC($\bar{q}_j\parallel g_k
\parallel g_1$) to $a_4^{0,d}(i_q,1_g,k_g,j_{\bar{q}})$. Terms that contribute
in both limits (i.e.\ those ones that contain $s_{1k}$ in the denominator)
remain unassigned at this stage.  This procedure already ensures the correct
assignment of $\mathrm{DC}(q_i\parallel g_1, \bar{q}_j\parallel g_k)$ and
$\mathrm{C}(q_i\parallel g_1)$ to $a_4^{0,c}(i_q,1_g,k_g,j_{\bar{q}})$.

In a second step, the simple collinear limits $\mathrm{C}(\bar{q}_j\parallel
g_k)$ and $\mathrm{C}({g}_1\parallel g_k)$ as well as the soft limit
$\mathrm{S}(k)$ are analysed by marking the respective progenitor terms in
$A_4^0(i_q,1_g,k_g,j_{\bar{q}})$ and assigning them to either $a_4^{0,c}$ or
$a_4^{0,d}$ (taking account of single unresolved behaviour of the previously
assigned triple-collinear terms), such that~\eqref{eq:a40singlelim} are
fulfilled. For simplicity, the limits are taken in all-final kinematics, but the
resulting decompositions are valid in any kinematics.  The limits
$\mathrm{C}(\bar{q}_j\parallel g_k)$ and $\mathrm{S}(k)$ are straightforward,
while $\mathrm{C}({g}_1\parallel g_k)$ is more involved due to the occurrence of
angular terms in the gluon-to-gluon splitting. In the implementation of the
antenna subtraction method, these terms are removed from matrix elements and
subtraction terms by appropriate averages over phase space points that are
related by angular rotations.  The decomposition into $a_4^{0,c}$ or $a_4^{0,d}$
must ensure that these averages still work at the level of the sub-antenna
functions.

The limit $\mathrm{C}({g}_1\parallel g_k)$ is taken using a Sudakov
parametrization of the momenta~\cite{Catani:1996vz}:
\begin{eqnarray}
p_1^\mu = z p^\mu + k_T^\mu - \frac{k_T^2}{z} \frac{1}{2p\cdot n} n^\mu\,, \nonumber \\
p_k^\mu = (1-z) p^\mu - k_T^\mu - \frac{k_T^2}{1-z} \frac{1}{2p\cdot n} n^\mu
\end{eqnarray}
with
 \begin{equation}
2p_1\cdot p_k = -\frac{k_T^2}{z(1-z)}\,, \quad p^2 = 0\,, \quad n^2=0\,, \quad p\cdot k_T = 0\,, \quad n\cdot k_T = 0\,.
\end{equation}
In this parametrization, $p^\mu$ is the composite momentum of the collinear
cluster, while $n^\mu$ is an arbitrary light-like direction.  The collinear
limit is then taken as Taylor expansion in $k_T^\mu$, retaining terms up to
second power, and performing the angular average in $d=4-2\epsilon$ dimensions
over the transverse direction of $k_T^\mu$ in the $(p,n)$ center-of-momentum
frame:
\begin{equation}
\langle k_T^\mu \rangle = 0\,, \quad \langle k_T^\mu k_T^\nu \rangle = \frac{k_T^2}{d-2} \left(g^{\mu\nu} - \frac{p^\mu n^\nu+p^\nu n^\mu}{p\cdot n}\right)\,. 
\end{equation}
The reference momentum $n^\mu$ is kept symbolic. The collinear
$\mathrm{C}({g}_1\parallel g_k)$ behaviour of the full antenna function
$A_4^0(i_q,1_g,k_g,j_{\bar{q}})$ is independent on $n^\mu$, but individual terms
extracted from it will display a dependence on $n^\mu$ in the collinear limit.
The terms are sorted into $a_4^{0,c}$ or $a_4^{0,d}$ in such a manner that both
sub-antenna functions remain independent on $n^\mu$ when taking the collinear
limit.

The decomposition into sub-antennae introduces polynomial denominators in the
invariants into $a_4^{0,c}$ and $a_4^{0,d}$. These are unproblematic at the
level of the unintegrated subtraction terms, but may pose an obstruction to
their analytical integration. However, when summing over all colour orderings
and by allowing for momentum relabelling of different phase-space mappings that
correspond to the same phase-space factorization (retaining of course the
correct identification of the identified charm quark in the reduced matrix
element), we can always combine $a_4^{0,c}$ and $a_4^{0,d}$ into a full $A_4^0$
at the level of the integrated subtraction term at VV level. Consequently, no
new integrated antenna functions are needed.

The subleading-colour $\tilde{A}_4^0(i_q,1_g,k_g,j_{\bar{q}})$ antenna function
and the $B_4^0(i_q,1_{q'},k_{\bar{q}'},j_{\bar{q}})$ antenna function containing
a secondary quark-antiquark pair were decomposed in the same way.  In addition,
the quark-antiquark one-loop antenna functions present at real-virtual level and
given in the final-final kinematics in~\cite{Gehrmann-DeRidder:2005btv} also
need to be decomposed into sub-antennae. The decomposition is however much
easier than for the four-parton antennae, as those capture only single
unresolved limits of the real-virtual matrix-elements.

\section{Results}
\label{sec:results}

\subsection{Numerical setup} 
\label{sec:numerics}

We consider a generic setup for Run 2 at $\sqrt{s}=13$~TeV.  In particular, the
following fiducial cuts for jets and charged leptons are applied:
\begin{eqnarray*}
&p_{T,\ell} > 27~\GeV,\quad  |y_{\ell}| < 2.5,\quad  p_{T,j} > 20~\GeV,\quad |\eta_{j}| < 2.5,\\
&E_{T,\mathrm{miss}} > 20~\GeV,\quad M_{T,W} > 45~\GeV,\quad 
\Delta R(j,\ell) > 0.4.
\end{eqnarray*}
The transverse mass of the $W$-boson is defined as
\begin{equation}
  M_{T,W} = \sqrt{2\,p_{T,\ell}\,E_{T,\mathrm{miss}}\,(1-\cos\Delta\phi_{\ell\nu})}\,.
\end{equation}
The jets are reconstructed with the anti-$k_T$ algorithm~\cite{Cacciari:2008gp}
with $R=0.4$.  The selection of $c$-jets is performed using the flavour dressing
procedure described in~\cite{Gauld:2022lem}.

We use the PDF4LHC21 Monte Carlo PDF set~\cite{PDF4LHCWorkingGroup:2022cjn},
with $\alpha_s(M_Z) = 0.118$ and $n_f^{\rm max} = 5$, where both the PDF and
$\alpha_s$ values are accessed via LHAPDF~\cite{Buckley:2014ana}.  For the
electroweak input parameters, the results are obtained in the $G_{\mu}$-scheme,
using a complex mass scheme for the unstable internal particles, and we adopt
the following values for the input parameters:
\begin{eqnarray*}
&M_{Z}^\mathrm{os} = 91.1876~\GeV, \quad
\Gamma_{Z}^\mathrm{os} = 2.4952~\GeV,\\ &M_{W}^\mathrm{os} = 80.379~\GeV,\quad
\Gamma_{W}^\mathrm{os} = 2.085~\GeV, \quad
G_\mu = 1.1663787 \times 10^{-5}~\GeV^{-2}.
\end{eqnarray*}
We further adopt a non-diagonal CKM matrix, thus allowing for all possible
charged-current interactions with massless quarks, with Wolfenstein parameters
$\lambda = 0.2265$, $A = 0.79$, $\bar{\rho} = 0.141$ and $\bar{\eta} =
0.357$~\cite{ParticleDataGroup:2020ssz}.

For differential distributions, the impact of missing higher-order corrections
is assessed using the conventional 7-point scale variation prescription: the
values of factorisation ($\mu_F$) and renormalisation ($\mu_R$) scales are
varied independently by a factor of two around the central scale $\mu_0 \equiv
E_{T,W}$, with the additional constraint that $\frac{1}{2} \leq \mu_F/\mu_R \leq
2$.  The transverse energy $E_{T,W}$ is defined as
\begin{equation}\label{eq:ETW}
  E_{T,W} = \sqrt{ M_{\ell\nu}^2 + p_{T,\ell\nu}^2 }\,,
\end{equation}
with $M_{\ell\nu}$ the invariant mass of the lepton-neutrino pair, and
$p_{T,\ell\nu} \equiv |\vec{p}_{T,\ell\nu}|$ the transverse momentum of the
lepton-neutrino system.

When considering theoretical predictions for the ratio of distributions, we
estimate the uncertainties in an uncorrelated way between the numerator and
denominator i.e.\ by considering
\begin{equation} \label{eq:ratio_uncertainty}
  R (\mu_R,\mu_F;\mu'_R,\mu'_F) = \frac{\sigma^{W^{+}+\cjet}(\mu_R,\mu_F)}{\sigma^{W^{-}+\cjet}(\mu'_R,\mu'_F)}\,,
\end{equation}
providing a total of 31-points when dropping the extreme variations in any pair
of scales.

Our default setup requires each event to have at least one $c$-jet ({\em
  inclusive} setup). We further apply the OS$-$SS subtraction: we separately
consider events where the lepton from the $W$-decay has the opposite sign (OS)
or the same sign (SS) of that of the \cjet, and then we take the difference of
the corresponding distributions (OS$-$SS).
In our fixed-order predictions, the sign of the \cjet is defined as the net sign
of all the flavoured particles (i.e.\ $c$-quarks) that are assigned to the jet
at the end of the flavour dressing procedure.
When more than one $c$-jet is present, the leading-$p_T$ $c$-jet is used to
define the OS$-$SS subtraction.

In order to study how predictions are affected by these requirements on the
number and relative sign of \cjets, in some of the plots below we study
variations of the setup.
In particular, we will further consider the {\em exclusive} setup i.e.\ we
require the presence of one and only one \cjet in each event (but we allow for
any number of flavourless jets).
We will also individually consider OS and SS events, and their sum OS+SS
i.e.\ by not applying any OS$-$SS subtraction.
In such cases, we will adopt the notation incl./excl. and
\mbox{OS$-$SS}/\mbox{OS+SS}/OS/SS, to denote a specific setup.
Where not indicated, we understand the default setup (OS$-$SS incl.).

Our results pass the usual checks routinely done in the context of a NNLOJET
calculation ({\em spike-tests}~\cite{NigelGlover:2010kwr} at real, real-virtual
and double-real level; cancellation of infrared poles at virtual, real-virtual
and double-virtual level; independence of the results from the technical cut at
real, real-virtual and double-real level).
The NNLO QCD corrections to $W+c$-jet production were computed previously
in~\cite{Czakon:2020coa,Czakon:2022khx}. These results were used in the recent
CMS study of $W+c$-jet production~\cite{CMS:2023aim} at 13~TeV.  We cross
checked our numbers for the fiducial cross section with Table 12
of~\cite{CMS:2023aim}, by performing dedicated computations for the CMS setup,
finding good agreement at all perturbative orders, and for the OS/SS/OS$-$SS
components separately.

\subsection{Fiducial cross sections}
\label{sec:fidxs}

In this Section, we present numbers for the fiducial cross section at different
orders and for different setups.
In Tables~\ref{tab:WPC} and~\ref{tab:WMC} we show results for the \WPC and \WMC
processes respectively. Results are organised by perturbative order (rows) and
setup (columns). Each row corresponds to the cross section at LO ($\sigma^{\rm
  LO}$), NLO ($\sigma^{\rm NLO}$) or NNLO ($\sigma^{\rm NNLO}$), or to the NLO
($\Delta\sigma^{\rm NLO}$) or NNLO ($\Delta\sigma^{\rm NLO}$) contribution to
the total cross section.
Each column corresponds to a particular setup, as explained in
Section~\ref{sec:numerics}: OS$-$SS incl., OS$-$SS excl., OS+SS incl., OS+SS
excl..
We further show the theory-uncertainty envelope associated to 7-point scale
variation, expressed as percentage of the reported central value. The
statistical Monte Carlo error on the calculation is indicated as an uncertainty
on the last digit.
In Table~\ref{tab:RAT}, we consider the ratio of fiducial cross sections for the
\WPC and \WMC processes,
\begin{equation}
\mathcal{R}^\pm_c = \frac{\sigma(\WPC)}{\sigma(\WMC)}\,.
\end{equation}
We show results for such a ratio at LO, NLO and NNLO (rows), in different setups
(columns).

\setlength{\tabcolsep}{0.8em} 
\renewcommand{\arraystretch}{1.33}
\begin{table}[t]
\begin{tabular}{lllll}
$W^+ + \cjet$   & \multicolumn{1}{c}{OS$-$SS incl.} & \multicolumn{1}{c}{OS$-$SS excl.} & \multicolumn{1}{c}{OS+SS incl.} & \multicolumn{1}{c}{OS+SS excl.} \\ \cline{1-5}
$\sigma^{\rm LO}$  &  $91.34(1)^{+ 11.7 \% }_{- 9.5 \% }$  &   $91.34(1)^{+ 11.7 \% }_{- 9.5 \% }$ &   $91.34(1)^{+ 11.7 \% }_{- 9.5 \% }$  &   $91.34(1)^{+ 11.7 \% }_{- 9.5 \% }$  \\ 
$\Delta\sigma^{\rm NLO}$ &  $30.45(4)$  &  $30.24(4)$ &  $39.23(4)$  &  $38.12(4)$  \\ 
$\sigma^{\rm NLO}$  &  $121.79(4)^{ +    5.6 \% } _{ -    5.4 \% }$  &  $121.58(4)^{ +    5.6 \% } _{ -    5.4 \% }$ &  $130.56(4)^{ +    6.9 \% } _{ -    6.3 \% }$  &  $129.46(4)^{ +    6.8 \% } _{ -    6.2 \% }$  \\ 
$\Delta\sigma^{\rm NNLO}$ & $-2.3(8)$  & $-2.7(7)$ &  $4.5(7)$  &   $3.2(7)$ \\ 
$\sigma^{\rm NNLO}$  &  $119.5(8)^{ +    0.4 \% } _{ -    1.8 \% }$  &  $119.0(7)^{ +    0.1\% } _{ -    1.6\% }$ &  $135.1(8)^{ +    1.2 \% } _{ -    1.9 \% }$  & $132.7(7)^{ +    0.6 \% } _{ -    1.5 \% }$  \\ 
\end{tabular}
\caption{\label{tab:WPC} Inclusive and exclusive fiducial cross sections for $\sigma(W^+ + \cjet)$ in OS$-$SS and OS+SS cases. We show the Monte Carlo errors as an uncertainty on the last digit while the percentage errors show the 7-point scale variation envelope.}
\end{table}

\setlength{\tabcolsep}{0.8em} 
\renewcommand{\arraystretch}{1.33}
\begin{table}[t]
\begin{tabular}{lllll}
$W^- + \cjet$   & \multicolumn{1}{c}{OS$-$SS incl.} & \multicolumn{1}{c}{OS$-$SS excl.} & \multicolumn{1}{c}{OS+SS incl.} & \multicolumn{1}{c}{OS+SS excl.} \\ \cline{1-5}
$\sigma^{\rm LO}$  &  95.782(4)$^{+ 11.7 \% }_{- 9.5 \% }$  &   95.782(4)$^{+ 11.7 \% }_{- 9.5 \% }$ &   95.782(4)$^{+ 11.7 \% }_{- 9.5 \% }$  &  95.782(4)$^{+ 11.7 \% }_{- 9.5 \% }$  \\ 
$\Delta\sigma^{\rm NLO}$  &  32.244(8)&  32.004(8) &  39.011(8) &  38.043(8) \\ 
$\sigma^{\rm NLO}$  &  128.026(9)$^{+ 5.7 \% }_{- 5.5 \% }$  &  127.786(9)$^{+ 5.7 \% }_{- 5.5 \% }$ &   134.794(9)$^{ +    6.6 \% } _{ -    6.1 \% }$   &  133.826(9)$^{ +    6.5 \% } _{ -    6.0 \% }$  \\ 
$\Delta\sigma^{\rm NNLO}$  & 2.9(5)  & 2.5(5) &  8.2(5)  &  7.1(5) \\ 
$\sigma^{\rm NNLO}$ &  130.9(5)$^{+ 0.7 \% }_{- 1.5 \% }$  &  130.3(5)$^{+ 0.9 \% }_{- 1.5 \% }$ &  143.0(5)$^{ +    1.5 \% } _{ -    2.5 \% }$  &  141.0(5)$^{ +    1.1 \% } _{ -    2.4 \% }$  \\ 
\end{tabular}
\caption{\label{tab:WMC} Inclusive and exclusive fiducial cross sections for $\sigma(W^- + \cjet)$ in OS$-$SS and OS+SS cases. As in Table~\ref{tab:WPC}, we show the Monte Carlo errors as an uncertainty on the last digit while the percentage errors show the 7-point scale variation envelope.}

\end{table}

\setlength{\tabcolsep}{0.75em} 
\renewcommand{\arraystretch}{1.2}
\begin{table}[t]
\begin{tabular}{lllll}
$\mathcal{R}^{\pm}_{c}$   & \multicolumn{1}{c}{OS$-$SS incl.} & \multicolumn{1}{c}{OS$-$SS excl.} & \multicolumn{1}{c}{OS+SS incl.} & \multicolumn{1}{c}{OS+SS excl.} \\ \cline{1-5}
LO  &  0.9536(3)$^{ +  23.4 \% }_{-  19.0\% }$   &  0.9536(3)$^{+  23.4 \% }_{-  19.0\%  }$  & 0.9536(3)$^{+  23.4 \% }_{-  19.0\%  }$  &  0.9536(3)$^{+  23.4 \% }_{-  19.0\%  }$   \\  
NLO  & 0.951 (1)$^{ +   13.9 \% } _{ -   12.1 \% }$   &  0.952(1)$^{ +   13.6 \% } _{ -   11.9 \% }$  &  0.968(1)$^{ +   13.9 \% } _{ -   12.1 \% }$  &  0.967(1)$^{ +   13.6 \% } _{ -   11.9 \% }$   \\  
NNLO  & 0.91(1)$^{ +    1.9 \% } _{ -    2.7 \% }$    &  0.91(1)$^{ +    1.6 \% } _{ -    2.5 \% }$&  0.94(1)$^{ +    3.7 \% } _{ -    3.3 \% }$  &  0.94(1)$^{ +    3.0 \% } _{ -    2.6 \% }$   \\  
\end{tabular}
\caption{\label{tab:RAT} Inclusive and exclusive fiducial cross sections for the ratio $\mathcal{R}^\pm_c =  \sigma(W^+ + \cjet)/\sigma (W^- + \cjet)$ in OS$-$SS and OS+SS cases. As in Table~\ref{tab:WPC}, we show the Monte Carlo errors as an uncertainty on the last digit while the percentage errors show the 7-point scale variation envelope.}

\end{table}

For both the individual processes \WPC and \WMC and for the ratio, we note
excellent perturbative convergence, with small NNLO corrections and a converging
pattern. The size of the theory uncertainty band progressively decreases when
moving from LO to NNLO, with an uncertainty of $\pm$10\% at LO, $\pm$5\% at NLO
and $\pm$1--2\% at NNLO for \WC. As for $\RAT$, the decrease in size is even
more pronounced, with an uncertainty of $\pm$20\% at LO, $\pm$10\% at NLO and
$\pm$2--3\% at NNLO.

Moving to the comparison of different setups, we notice interesting hierarchies
between the numbers in the Tables. At LO, the fiducial cross section is always
the same regardless of the setup, due to the presence of a single OS charm quark
in the final state. When moving to NLO or NNLO, thus allowing for the presence
of more charm quarks or anti-quarks in the event, the size of the difference
between OS+SS and OS$-$SS increases, with a larger difference at NNLO than at
NLO, and in \WPC than in \WMC.
The difference between the inclusive and exclusive setup is more moderate, with
numbers usually compatible within the scale variation uncertainties, and with a
larger difference at NNLO than at NLO. The latter observation could be explained
by the fact that the probability of having two or more \cjets in the event is
small, where there are at most 2 and 3 charm (anti-)quarks in the event at NLO
and at NNLO respectively.
Similar comments apply to $\RAT$.

Finally, we note that the values of $\RAT$ in Table~\ref{tab:RAT} are all
smaller than 1, whatever the perturbative order and the setup i.e.\ the fiducial
cross section for \WPC is always (slightly) smaller than the fiducial cross
section for \WMC. This fact can be explained by an analysis of the couplings
allowed by the CKM matrix and the behaviours of the parton distribution
functions of the proton. At LO, the size of the contribution proportional to
$|V_{cs}|$ is equivalent for $W^{+}+\bar{c}$ and $W^{-}+{c}$, because the
strange and anti-strange PDFs are similar. However, the subleading contribution
proportional to $|V_{ds}|$ is different between $W^{+}+\bar{c}$ and $W^{-}+{c}$:
namely, the down PDF contributing to $W^{-}+{c}$ features a valence component,
which is missing in the anti-down PDF contributing to $W^{+}+\bar{c}$. Hence,
the cross section for \WMC is larger than for \WPC at LO, and higher-order
corrections are not large enough to alter this simple picture.
This insight will be instrumental in explaining differences in behaviour between
the differential distributions for \WPC and \WMC shown in
Section~\ref{sec:distrs}, and will be further explored in
Section~\ref{sec:chbreak}, where the contributions of individual partonic
channels to the total cross section will be presented.

\subsection{Differential distributions}
\label{sec:distrs}

\begin{figure}[tp]
  \centering
  \includegraphics[width=0.47\textwidth]{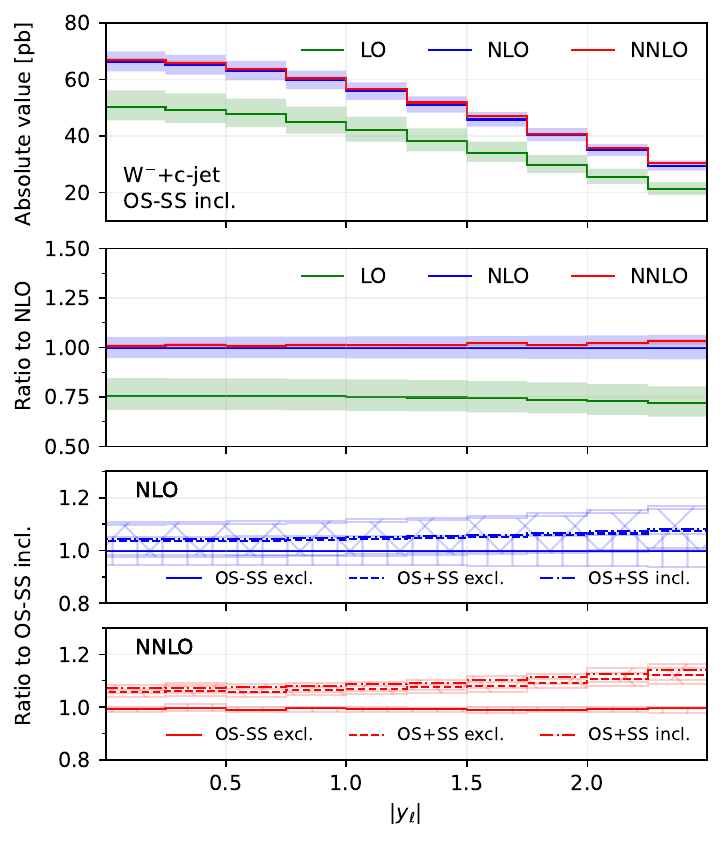}
  \includegraphics[width=0.47\textwidth]{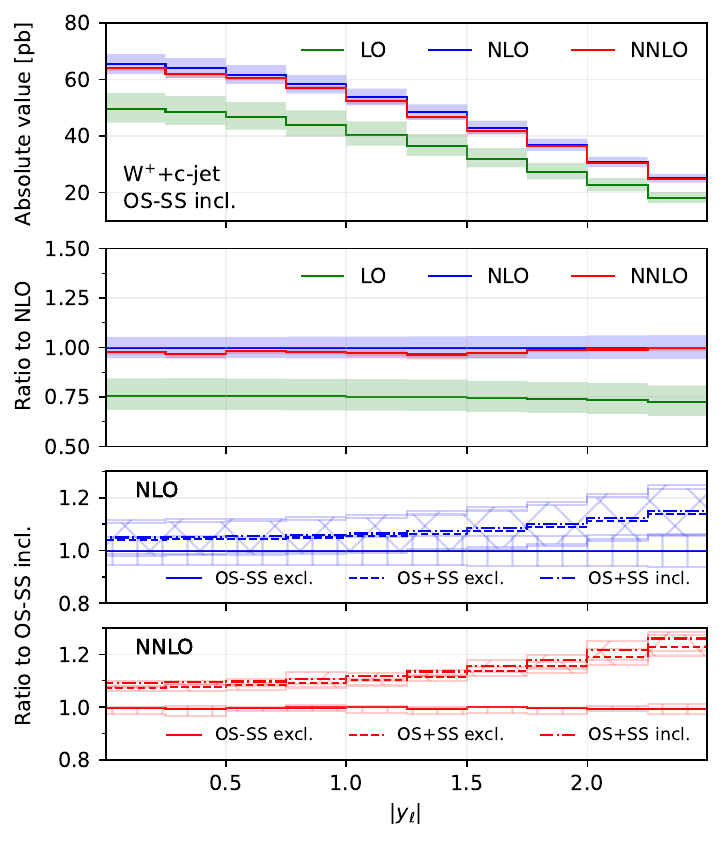}    
  \caption{Comparison of predictions for the absolute rapidity of the lepton
    $|y_{\ell}|$, in the \WMC (left) and \WPC (right) processes.  Panels from
    top to bottom: differential distribution at different orders; ratio of
    differential distributions to NLO; ratio of (OS$-$SS, excl.), (OS+SS, excl.)
    and (OS+SS, incl.) distributions to (OS$-$SS, incl.) at NLO; same for NNLO.}
  \label{fig:DISTR_abs_yl}
\end{figure}

\begin{figure}[tp]
  \centering  
  \includegraphics[width=0.47\textwidth]{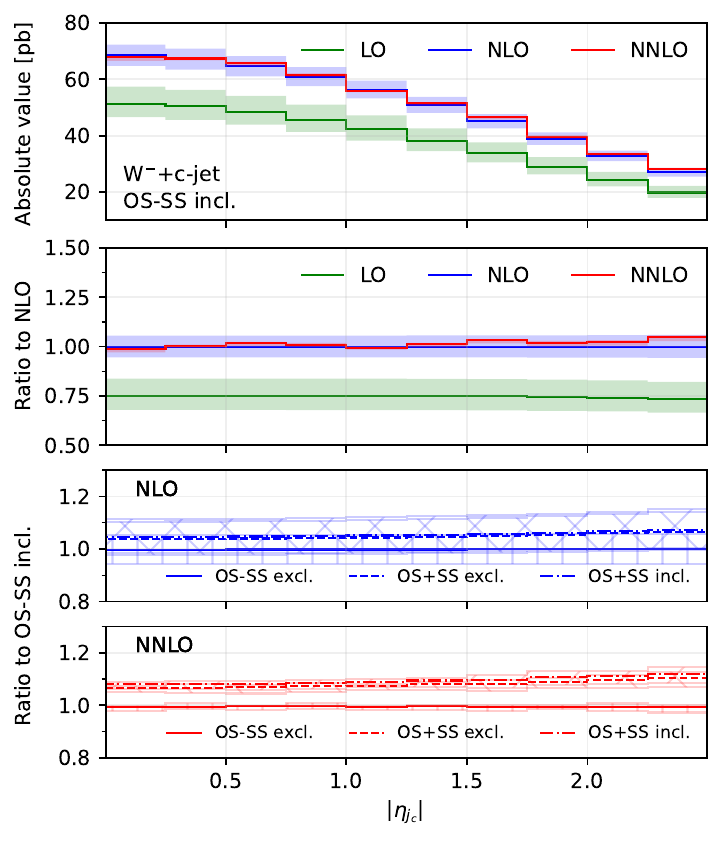}
  \includegraphics[width=0.47\textwidth]{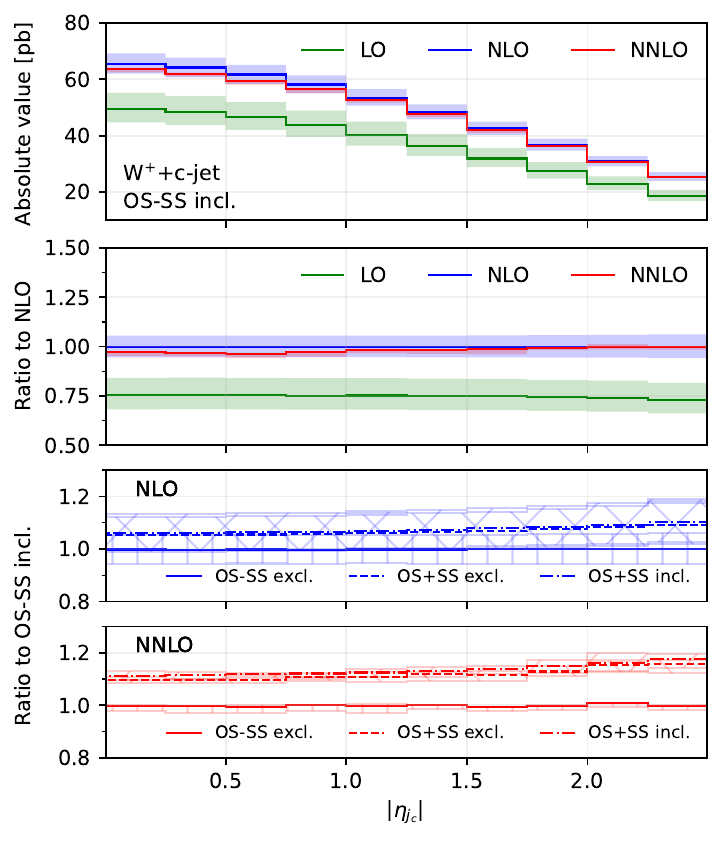}
  \caption{Comparison of predictions for the absolute pseudorapidity of the
    leading \cjet \etac, in the \WMC (left) and \WPC (right) processes.  Panels
    from top to bottom: differential distribution at different orders; ratio of
    differential distributions to NLO; ratio of (OS$-$SS, excl.), (OS+SS, excl.)
    and (OS+SS, incl.) distributions to (OS$-$SS, incl.) at NLO; same for NNLO.}
    \label{fig:DISTR_flav_abs_etaj1}
\end{figure}

\begin{figure}[tp]
  \centering  
  \includegraphics[width=0.47\textwidth]{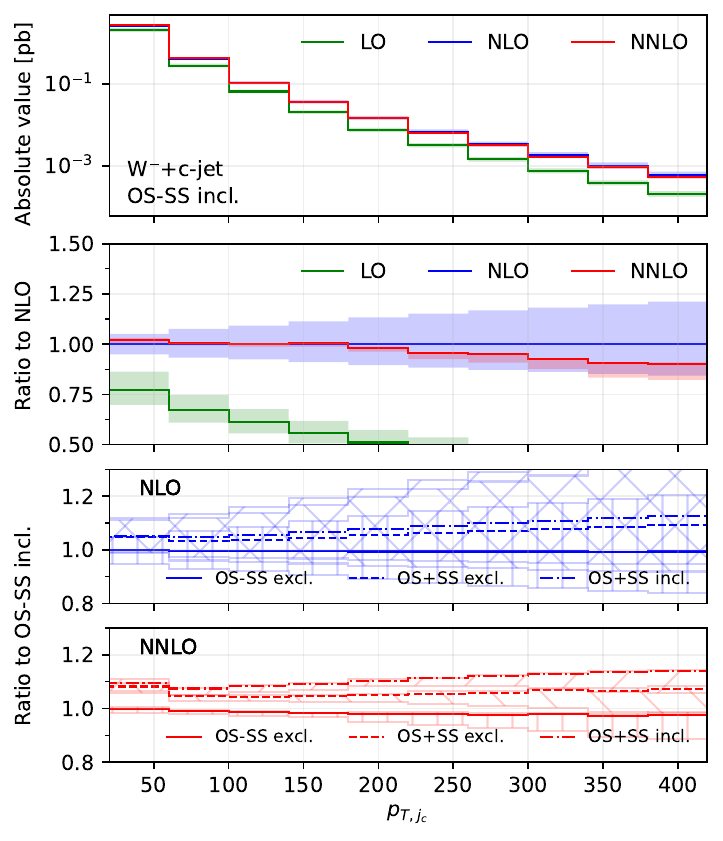}
  \includegraphics[width=0.47\textwidth]{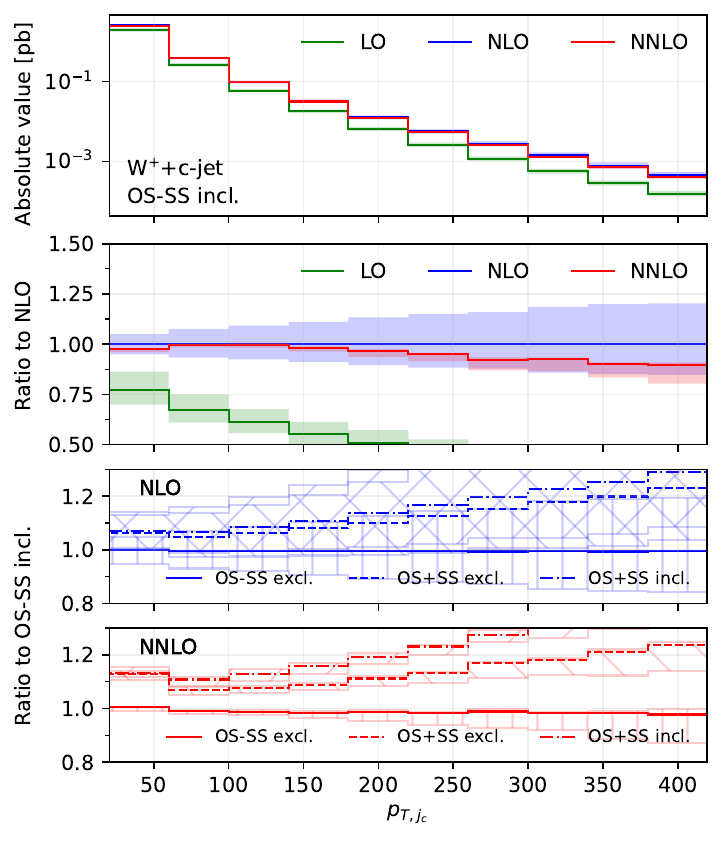}    
  \caption{Comparison of predictions for the transverse momentum of the leading
    \cjet \ptc, in the \WMC (left) and \WPC (right) processes.  Panels from top
    to bottom: differential distribution at different orders; ratio of
    differential distributions to NLO; ratio of (OS$-$SS, excl.), (OS+SS, excl.)
    and (OS+SS, incl.) distributions to (OS$-$SS, incl.) at NLO; same for NNLO.}
  \label{fig:DISTR_flav_ptj1}
\end{figure}

\begin{figure}[tp]  
  \centering  
  \includegraphics[width=0.47\textwidth]{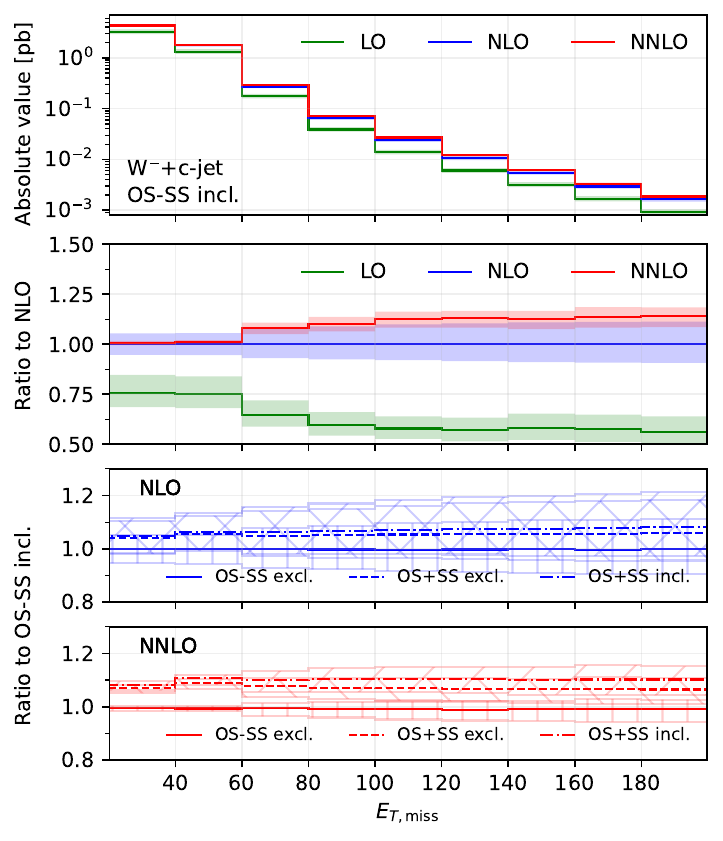}
  \includegraphics[width=0.47\textwidth]{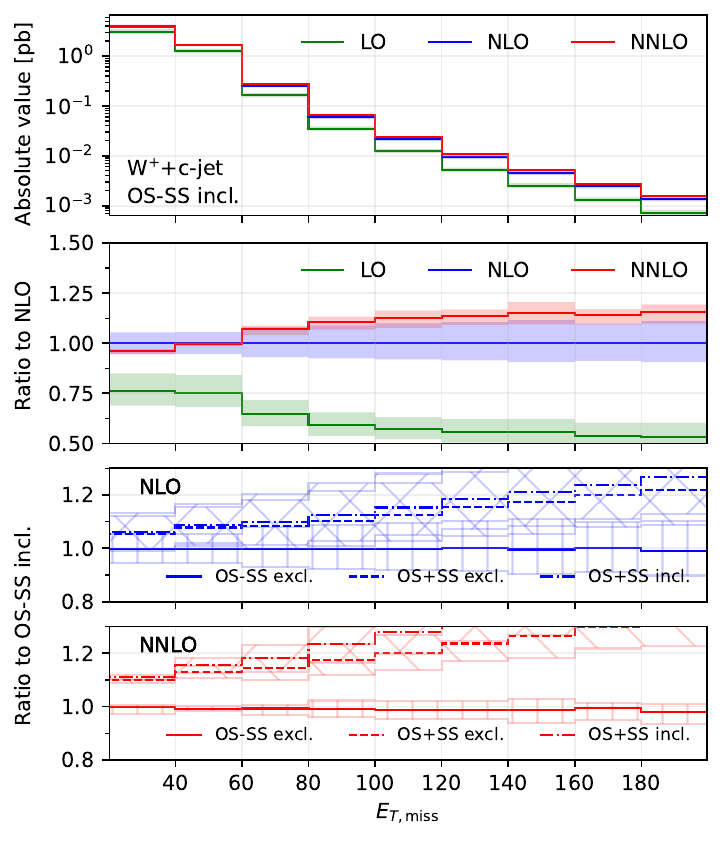}    
  \caption{Comparison of predictions for the transverse missing energy \etmiss,
    in the \WMC (left) and \WPC (right) processes.  Panels from top to bottom:
    differential distribution at different orders; ratio of differential
    distributions to NLO; ratio of (OS$-$SS, excl.), (OS+SS, excl.) and (OS+SS,
    incl.) distributions to (OS$-$SS, incl.) at NLO; same for NNLO.}
    \label{fig:DISTR_etmiss}
\end{figure}

\begin{figure}[tp]
  \centering  
  \includegraphics[width=0.47\textwidth]{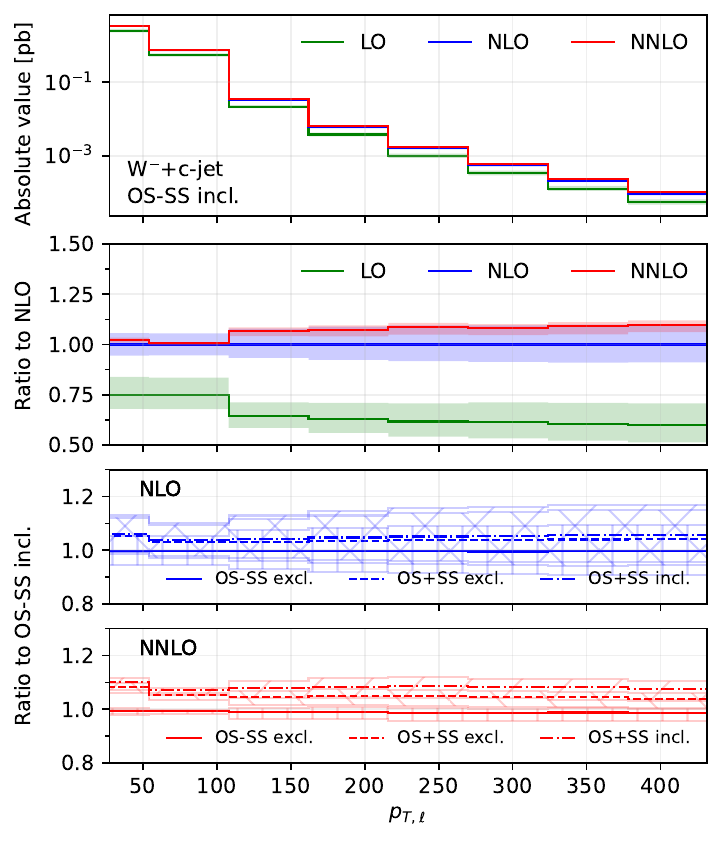}
  \includegraphics[width=0.47\textwidth]{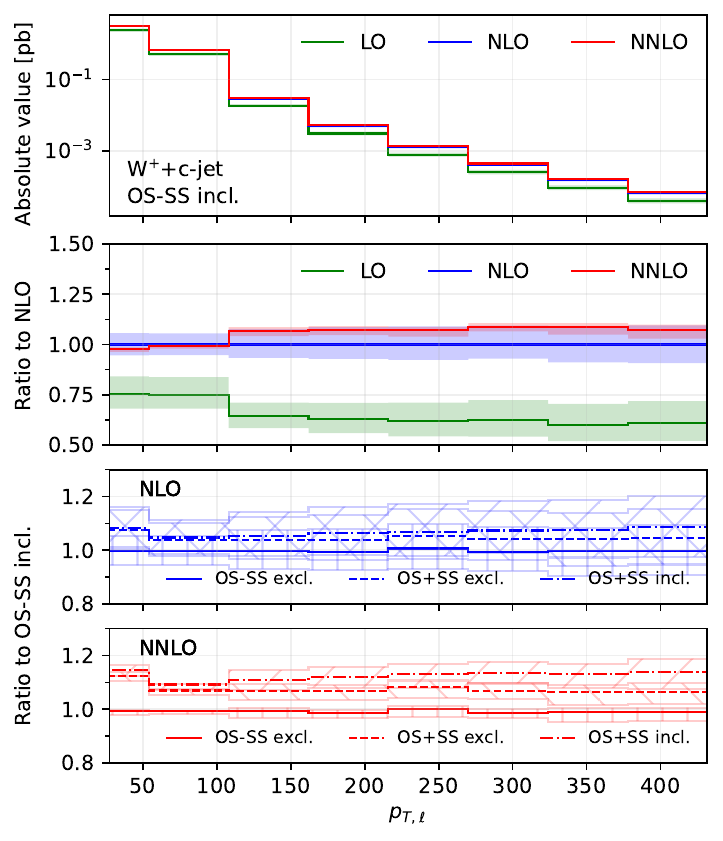}
  \caption{Comparison of predictions for the transverse momentum of the lepton
    \ptl, in the \WMC (left) and \WPC (right) processes.  Panels from top to
    bottom: differential distribution at different orders; ratio of differential
    distributions to NLO; ratio of (OS$-$SS, excl.), (OS+SS, excl.) and (OS+SS,
    incl.) distributions to (OS$-$SS, incl.) at NLO; same for NNLO.}
  \label{fig:DISTR_ptl}
\end{figure}

\begin{figure}[tp]
  \centering  
  \includegraphics[width=0.47\textwidth]{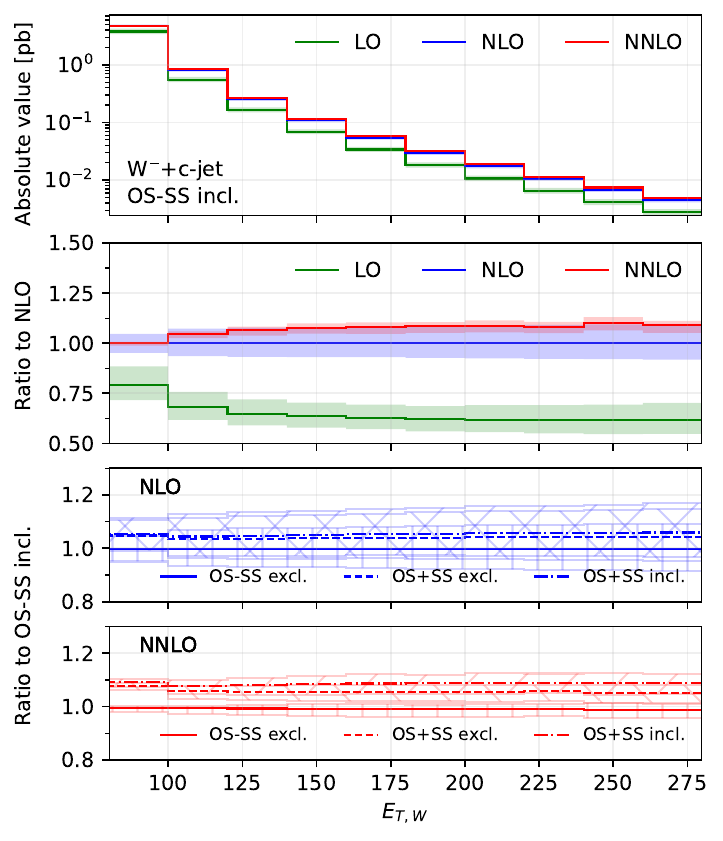}
  \includegraphics[width=0.47\textwidth]{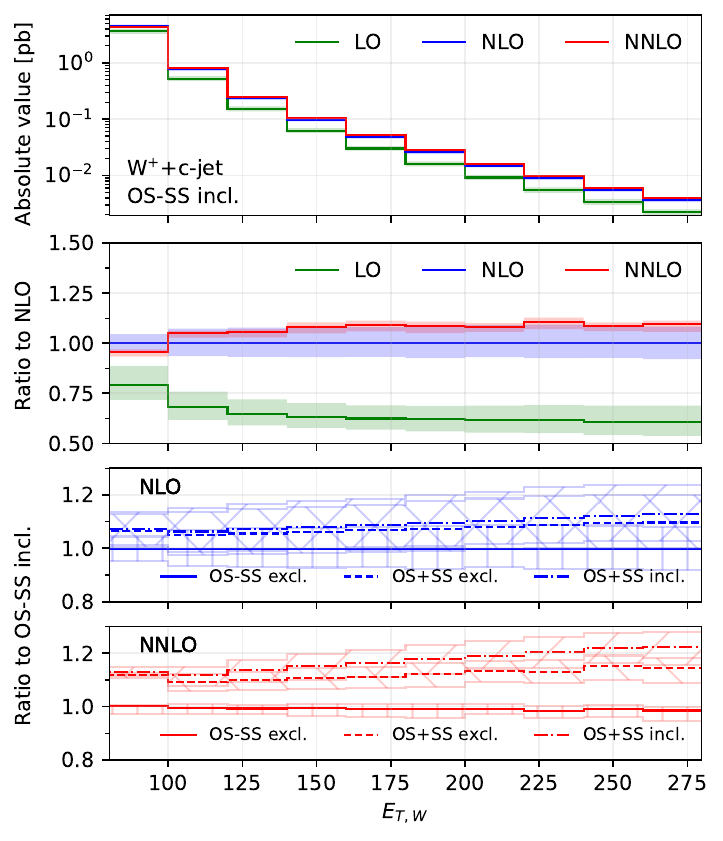}
  \caption{Comparison of predictions for the transverse energy \etw, in the \WMC
    (left) and \WPC (right) processes. Panels from top to bottom: differential
    distribution at different orders; ratio of differential distributions to
    NLO; ratio of (OS$-$SS, excl.), (OS+SS, excl.) and (OS+SS, incl.)
    distributions to (OS$-$SS, incl.) at NLO; same for NNLO.}
  \label{fig:DISTR_etw}
\end{figure}

In this Section we present differential distributions for several observables of
interest, for both the \WPC and \WMC process.
We consider the absolute rapidity of the lepton from the $W^{\pm}$ decay, \yl
(Figure~\ref{fig:DISTR_abs_yl}), the absolute pseudorapidity of the
leading-$p_T$ \cjet, \etac (Figure~\ref{fig:DISTR_flav_abs_etaj1}), the
transverse momentum of the leading-$p_T$ \cjet, \ptc
(Figure~\ref{fig:DISTR_flav_ptj1}), the transverse missing energy, \etmiss
(Figure~\ref{fig:DISTR_etmiss}), the transverse momentum of the lepton from the
$W^{\pm}$ decay, \ptl (Figure~\ref{fig:DISTR_ptl}) and the transverse energy
\etw defined as in~\eqref{eq:ETW} (Figure~\ref{fig:DISTR_etw}).

Figures~\ref{fig:DISTR_abs_yl}--\ref{fig:DISTR_etw} are organised in the
following way.  On the left we show distributions for \WMC, on the right for
\WPC.  Each column has four panels, depicting: absolute value of the
differential distribution at LO, NLO and NNLO in the OS$-$SS incl. setup
(1$^{\rm st}$ panel from the top); ratio of distributions in the OS$-$SS
incl. setup at LO, NLO, NNLO to NLO prediction (2$^{\rm nd}$ panel from the
top); ratio of OS$-$SS excl., OS+SS excl.\ and OS+SS incl.\ distributions to
OS$-$SS incl.\ distribution at NLO (3$^{\rm rd}$ panel from the top) and at NNLO
(4$^{\rm th}$ panel from the top).

\begin{figure}[tp]
  \centering  
  \includegraphics[width=0.32\textwidth]{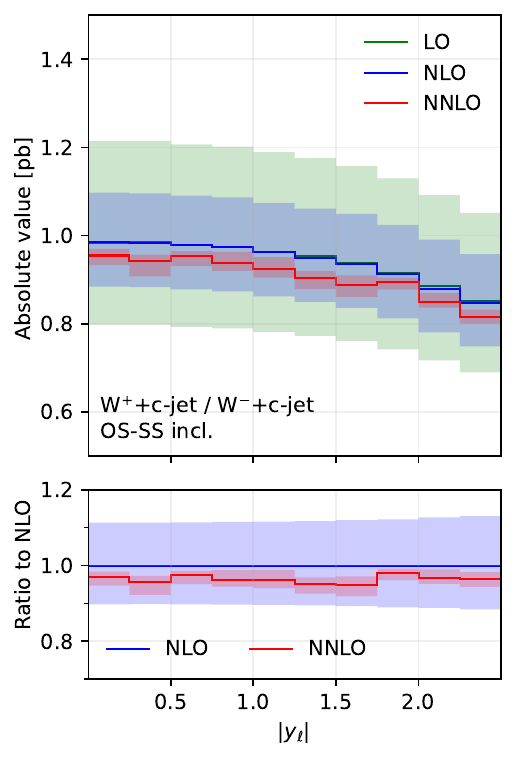}
  \includegraphics[width=0.32\textwidth]{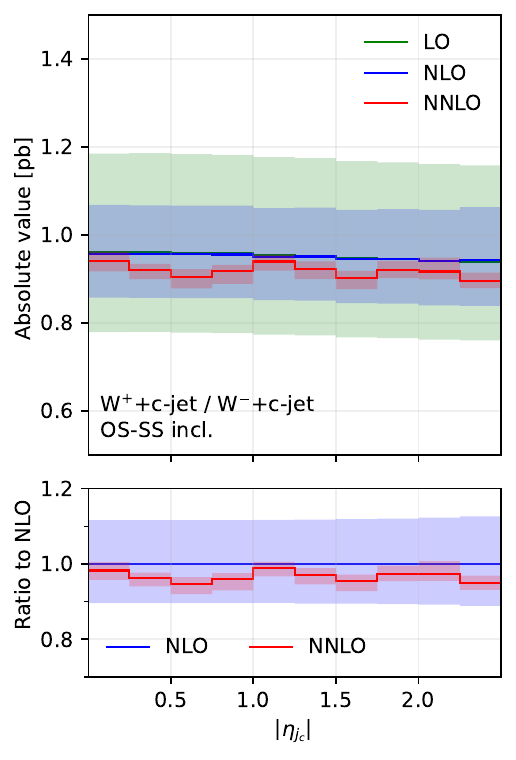}
  \includegraphics[width=0.32\textwidth]{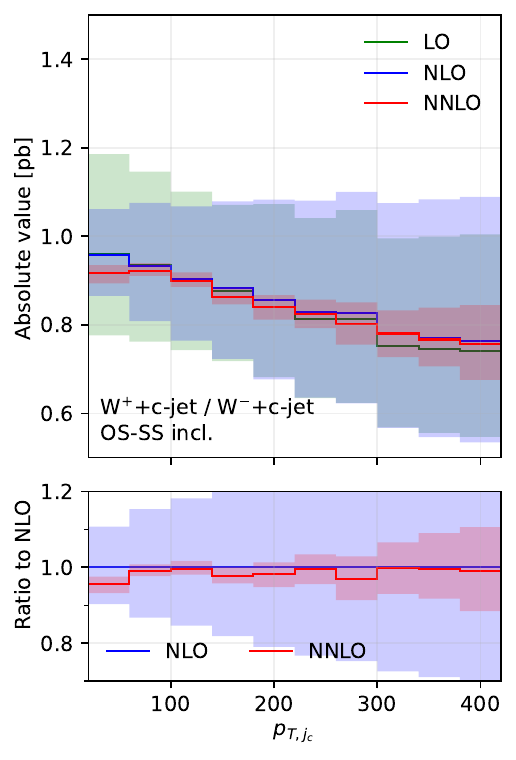}      
  \caption{Comparison of differential distributions at different orders for the
    ratio $\sigma$(\WPC)/$\sigma$(\WMC), differential in \yl (left), \etac
    (middle), \ptc (right). The upper panels show the distributions in absolute
    value, whereas the lower panels show the ratio to the NLO prediction.}
  \label{fig:RATIO}
\end{figure}

Finally, in Figure~\ref{fig:RATIO} we show the distributions differential in \yl
(first column), \etac (second column) and \ptc (third column), by considering
both distributions in absolute value at LO, NLO and NNLO (upper panels) and
their ratio to the NLO prediction (lower panels). Here all the predictions are
in the OS$-$SS incl.\ setup.

We first focus on the OS$-$SS incl.\ setup and we consider predictions at
different perturbative orders.  We observe in all of the
Figures~\ref{fig:DISTR_abs_yl}--\ref{fig:RATIO} a nice perturbative convergence,
with the NNLO curves contained within the NLO uncertainty bands, and with the
NNLO uncertainty band always smaller by at least a factor of two compared to the
NLO one.  In the \ptc, \etmiss, \ptl and \etw distributions, the NNLO curve lies
just on the boundary of the NLO uncertainty band.  For the ratio \RAT in
Figure~\ref{fig:RATIO}, we observe a drastic reduction of the theory uncertainty
when moving from LO to NNLO for all the considered distributions, in line with
what is observed for the ratio of fiducial cross sections in
Section~\ref{sec:fidxs}.

By focussing now on the comparison between different setups, we can draw similar
conclusions to those already expressed in Section~\ref{sec:fidxs}.  Namely: the
difference between excl.\ and incl.\ is greater at NNLO than at NLO (remember
that at LO all the setups are the same); the difference between excl.\ and
incl.\ is greater in the OS+SS case rather than in the OS$-$SS case; the
difference between OS$-$SS and OS+SS is generally larger than the difference
between excl.\ and incl.  However, such differences are generally not flat in
the differential distributions.  While we observe that the differences between
setups mildly depend on \etac, \ptl and \etw for both \WPC and \WMC, we note a
significant dependence on \yl, \ptc and \etmiss. In particular, such a
dependence is more pronounced at large values of \yl, \ptc and \etmiss, and the
behaviour of \WMC and \WPC is very different, with enhanced differences for \WPC
between the different set-ups.  We will return to this point in
Section~\ref{sec:chbreak} below.

We conclude this Section by observing that the difference between excl.\ and
incl.\ in the OS$-$SS case is very small in all the distributions, both at NLO
and NNLO: it amounts to at most a couple of per-cent for high values of \ptc.
The OS$-$SS subtraction clearly helps in reducing the difference between the
inclusive and exclusive prescription on the number of \cjets, because the events
discarded when performing the OS$-$SS subtraction are a subset of events with
more than one $c$ parton in the event.
However, it seems that OS$-$SS subtraction is very efficient in discarding
events with more than one \cjet surviving the fiducial cuts.
In other words, the inclusive two \cjets cross section is very small when
applying the OS$-$SS subtraction.

\section{Partonic channel breakdown}
\label{sec:chbreak}

In this section, we study how the individual partonic channels contribute to the
total cross section.
This analysis will be instrumental in understanding how higher-order radiative
corrections in different setups affect the contributions coming from different
PDFs.
\begin{figure}[t]
  \centering  
  \includegraphics[width=0.49\textwidth]{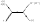}  
  \caption{Born-level diagram for  \WPC and \WMC.}
  \label{fig:born}
\end{figure}
\begin{figure}[t]
  \centering  
  \includegraphics[width=0.42\textwidth]{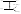}  
  \includegraphics[width=0.49\textwidth]{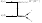}
  \caption{Example diagrams contributing to \WPC and \WMC from NLO onwards.}
  \label{fig:diags}
\end{figure}

We recall that at LO, $W+c$-jet production is mediated only through the
Born-level process $sg \to W^-c$ and $\bar{s}g \to W^+\bar{c}$
(Figure~\ref{fig:born}) and their CKM-suppressed $d$-quark initiated partner
processes. They always result in OS final states. At higher orders, final states
containing charm quarks can also be caused by a hard scattering process
involving an initial-state charm quark or by the splitting of a final state
gluon into a charm-anticharm pair, illustrated in Figure~\ref{fig:diags}.

In Tables~\ref{tab:CHB_WMC} and~\ref{tab:CHB_WPC} we present the contribution of
each partonic channel in the \WMC and in the \WPC process, respectively.
We provide numbers for OS at LO, NLO and NNLO, and for SS at NLO and NNLO (SS at
LO is trivially zero). One can easily obtain the corresponding numbers for
OS$-$SS and OS+SS.
All the numbers refer to exclusive cross sections; the analogous numbers for
inclusive cross sections are very similar, so throughout this section we will
focus on the exclusive setup (which is more easily interpreted in terms of
parton-level subprocesses), unless otherwise specified.

We have chosen to organise the partonic channels in the following way: we
explicitly distinguish charm $c$($\bar{c}$) and strange $s$($\bar{s}$)
(anti)quarks in the initial state, while denoting an (anti)quark of any other
flavour as $q$($\bar{q}$).
We do not differentiate between quarks and antiquarks i.e.\ we sum together
contributions coming from quarks and antiquarks of the same flavour.
In this way, we obtain 10 possible channels, as listed in the first column of
Tables~\ref{tab:CHB_WMC} and \ref{tab:CHB_WPC}, whose contributions sum up to
the total cross section.

\setlength{\tabcolsep}{0.65em} 
\renewcommand{\arraystretch}{1.05}
\begin{table}[t]
\centering
\begin{tabular}{c|d{5.4}d{5.4}d{5.4}d{5.4}d{5.4}}
\toprule
\WMC                &  \multicolumn{1}{r}{OS LO}    &   \multicolumn{1}{r}{OS NLO}    &  \multicolumn{1}{r}{SS NLO}   &  \multicolumn{1}{r}{OS NNLO}    & \multicolumn{1}{r}{SS NNLO}  \\
\midrule
$c(\bar c)s(\bar s)$  &    0.0          &   -0.1225(3) &    0.4852(2)     & -0.05(2)   &   0.842(3)   \\  
$c(\bar c)c(\bar c)$  &     0.0         &   0.2158(1)  & 0.2062(2)        & 0.360(2)   &  0.351(1)    \\  
$c(\bar c)q(\bar q)$  &     0.0         & 1.2392(3)    &  1.3132(4)       &  1.958(4)  & 2.088(4)     \\  
$s(\bar s)q(\bar q)$  &     0.0         & -0.651(3)    &   0.03134(1)     & -1.1(2)    & 0.0537(2)    \\  
$s(\bar s)s(\bar s)$  &     0.0         & -0.2549(3)   &    0.0           & -0.42(3)   &  0.0         \\  
$q(\bar q)q(\bar q)$  &     0.0         & 1.0314(7)    &    0.9838(4)     & 1.73(2)    & 1.676(6)     \\  
$gq(\bar q)$          &     8.9255(6)   & 12.700(1)    &    0.0           & 12.7(2)    & 0.405(3)     \\  
$gs(\bar s)$          &    86.857(4)    & 123.002(8)   &   0.0            & 128.9(3)   & -0.0353(6)   \\  
$gc(\bar c)$          &     0.0         & 0.0          &  0.0             &  -0.14 (2) & -0.057 (2)   \\  
$gg$                  &     0.0         &   -6.355(3)  &  0.0             & -8.31 (1)  &  0.0         \\
\midrule
total                 &    95.782(5)    &  130.806(1)  &  3.020(1)        &  135.6(5)    &   5.324 (9)       
\end{tabular}
\caption{\label{tab:CHB_WMC}%
Breakdown of the fiducial cross section for \WMC in terms of the contributing partonic channels. We denote as $q (\bar q)$  the quarks (antiquarks) of different flavour than $s (\bar s)$ and $c (\bar c)$. Furthermore, we do not distinguish between quarks and antiquarks e.g.\ the $c(\bar c)s(\bar s) $ row contains all the possible permutations of $c$ and $\bar c$ with $s$ and $\bar s$. All the numbers refer to exclusive cross sections.}
\end{table}

\setlength{\tabcolsep}{0.65em} 
\renewcommand{\arraystretch}{1.05}
\begin{table}[t]
\centering
\begin{tabular}{c|d{5.4}d{5.4}d{5.4}d{5.4}d{5.4}}
\toprule
\WPC                &  \multicolumn{1}{r}{OS LO}    &   \multicolumn{1}{r}{OS NLO}    &  \multicolumn{1}{r}{SS NLO}   &  \multicolumn{1}{r}{OS NNLO}    & \multicolumn{1}{r}{SS NNLO}  \\
\midrule
$c(\bar c)s(\bar s)$  &   0.0       &   -0.1191(9) &  0.4752(4)   &    -0.13(2)   &  0.838(1)   \\
$c(\bar c)c(\bar c)$  &   0.0       &    0.2151(3) &  0.2047(3)   &     0.3316(5) &  0.3246(6)  \\
$c(\bar c)q(\bar q)$  &   0.0       &    1.948(3)  &  1.988(4)    &     2.945(6)  &  3.038(6)   \\
$s(\bar s)q(\bar q)$  &   0.0       &   -0.649(9)  &  0.0673(1)   &    -1.9(3)    &  0.1157(3)  \\
$s(\bar s)s(\bar s)$  &   0.0       &   -0.258(1)  &  0.0         &    -0.55(5)   &  0.0        \\
$q(\bar q)q(\bar q)$  &   0.0       &    1.431(2)  &  1.409(2)    &     2.35(2)   &  2.423(6)   \\
$gq(\bar q)$          &   5.8299(7) &    8.257(2)  &  0.0         &    10.1(4)    &  0.508(4)   \\
$gs(\bar s)$          &  85.51(1)   &  121.04(3)   &  0.0         &   126.3(6)    & -0.0430(4)  \\
$gc(\bar c)$          &   0.0       &    0.0       &  0.0         &     0.02(2)   & -0.0293(7)  \\
$gg$                  &   0.0       &   -6.34(1)   &  0.0         &   -13.62(6)   &  0.0        \\
\midrule
total                 &  91.34(1)   &  125.51(4)   &  4.146(4)    &   125.9(7)      &   7.17(1)  
\end{tabular}
\caption{\label{tab:CHB_WPC}%
Breakdown of the fiducial cross section for \WPC in terms of the contributing partonic channels. As in Table~\ref{tab:CHB_WMC} we denote as $q (\bar q)$  the quarks (antiquarks) of different flavour than $s (\bar s)$ and $c (\bar c)$. Furthermore, we do not distinguish between quarks and antiquarks e.g.\ the $c(\bar c)s(\bar s) $ row contains all the possible permutations of $c$ and $\bar c$ with $s$ and $\bar s$. All the numbers refer to exclusive cross sections.}

\end{table}

At all perturbative orders, the by far dominant contribution to the fiducial
cross section in OS events comes from the $gs(\sb)$ channel, which amounts to
90\% of the total.
The second largest contribution (6--10\%) to OS events comes from the $gq(\qb)$
channel. Such a contribution is slightly larger for \WMC: as already explained
in Section~\ref{sec:fidxs}, this is related to the presence of the $d$ PDF in
\WMC as opposed to the presence of $\bar{d}$ in \WPC.
The third largest contribution (5--10\%) comes from the $gg$ channel, with a
negative sign, partially compensating the $gq(\qb)$ contribution.
In some cases, the $gg$ channel can be even larger than the $gq(\qb)$ one (for
instance in \WPC for OS at NNLO).
All the other channels contribute much less to the total cross section (at most
a few per-cent each).

It is interesting to compare the OS numbers for some channels with the analogous
ones for SS.
We notice that both at NLO and at NNLO, both for \WPC and for \WMC, the
$c(\cb)c(\cb)$ channel, the $c(\cb)q(\qb)$ channel and the $q(\qb)q(\qb)$
channel are numerically very similar between OS and SS.
Hence when performing the OS$-$SS subtraction, we are enhancing the channels
featuring a (anti)strange PDF, by removing channels with quarks of other
flavours.
The channels with a gluon PDF ($gq(\qb)$ and $gg$) still survive after the
OS$-$SS subtraction.

\begin{figure}[t]
  \centering
  \includegraphics[width=0.86\textwidth]{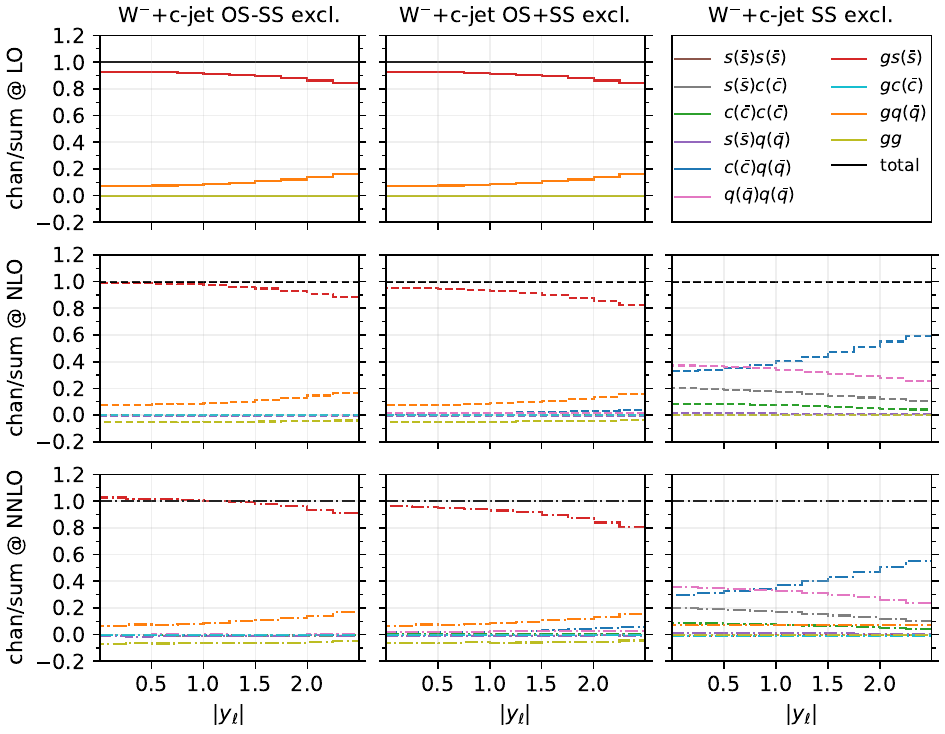}
  \caption{Fractional contribution of partonic channels to the total result at
    different perturbative orders, for the \WMC process, differential in
    $|y_{\ell}|$. The three columns correspond to different setups: OS$-$SS
    excl.\ (left), OS+SS excl.\ (middle), SS excl.\ (right). The three rows
    correspond to different perturbative orders: LO (top), NLO (middle), NNLO
    (bottom).}  \label{fig:CHB_M_yl}
\end{figure}

\begin{figure}[ht]
  \centering  
  \includegraphics[width=0.86\textwidth]{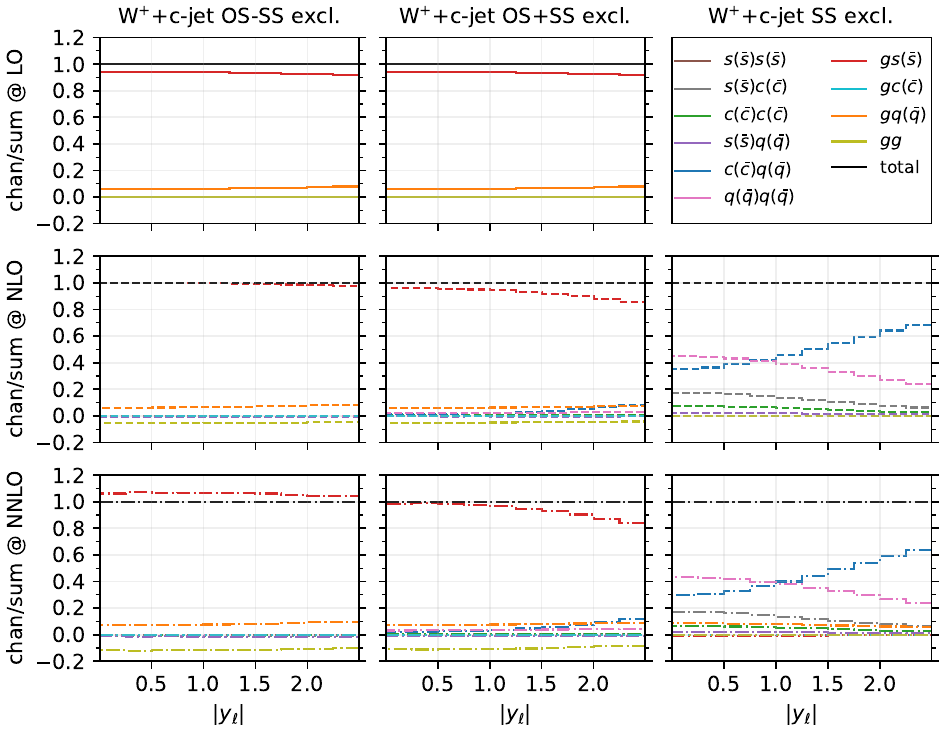}
  \caption{Fractional contribution of partonic channels to the total result at
    different perturbative orders, for the \WPC process, differential in
    $|y_{\ell}|$. The three columns correspond to different setups: OS$-$SS
    excl.\ (left), OS+SS excl.\ (middle), SS excl.\ (right). The three rows
    correspond to different perturbative orders: LO (top), NLO (middle), NNLO
    (bottom).}
  \label{fig:CHB_P_yl}
\end{figure}

\begin{figure}[ht]
  \centering  
  \includegraphics[width=0.86\textwidth]{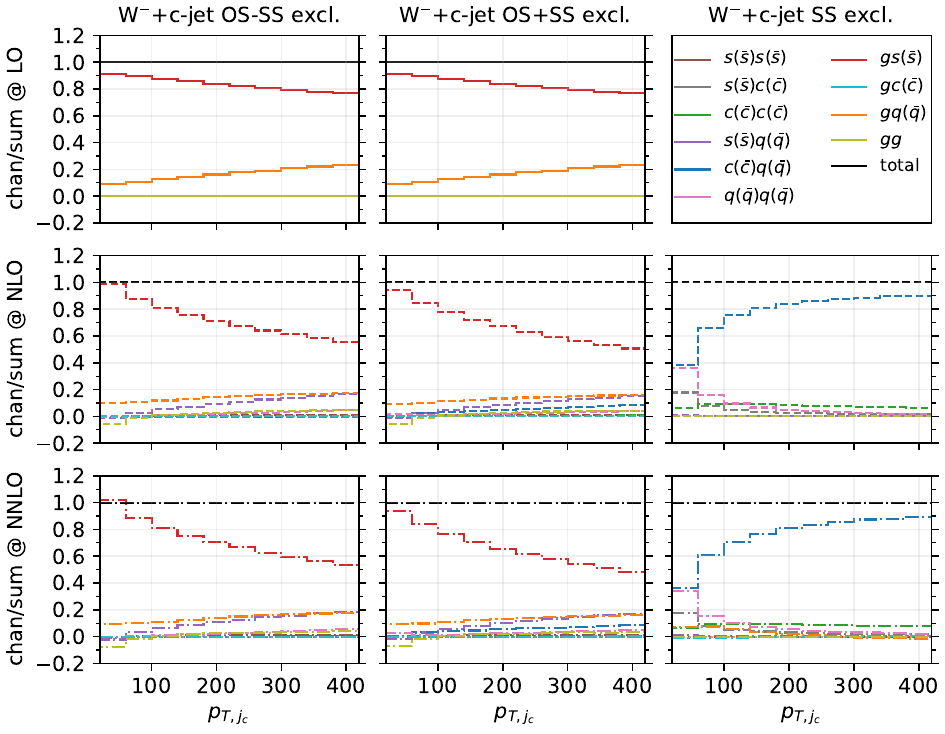}
  \caption{Fractional contribution of partonic channels to the total result at
    different perturbative orders, for the \WMC process, differential in
    $p_{T,j_c}$. The three columns correspond to different setups: OS$-$SS
    excl.\ (left), OS+SS excl.\ (middle), SS excl.\ (right). The three rows
    correspond to different perturbative orders: LO (top), NLO (middle), NNLO
    (bottom).}
  \label{fig:CHB_M_ptc}
\end{figure}

\begin{figure}[ht]
  \centering  
  \includegraphics[width=0.86\textwidth]{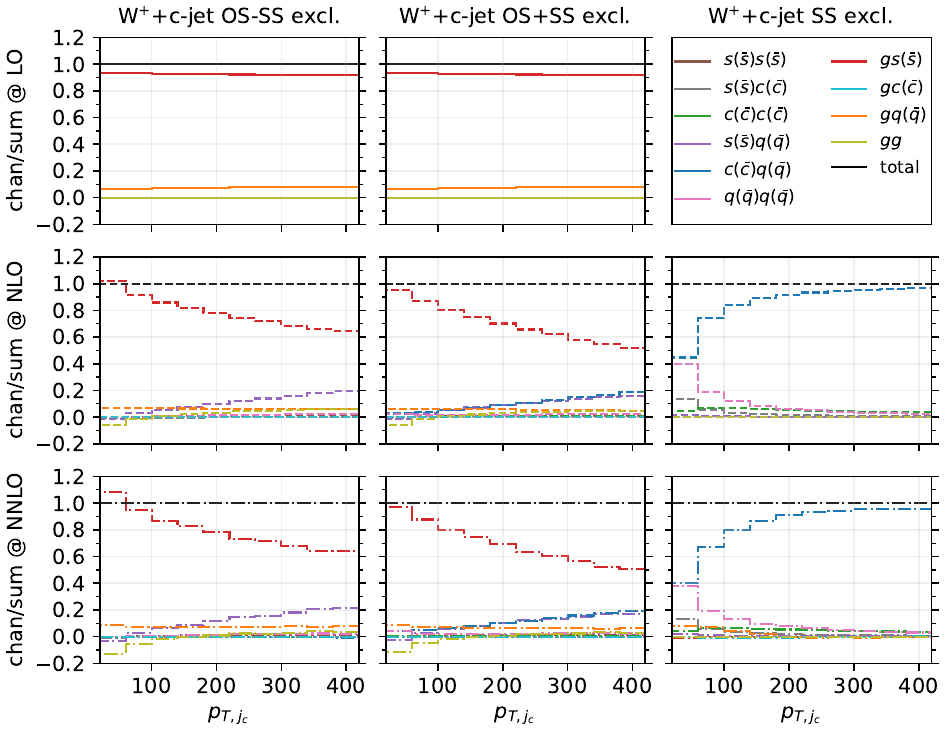}
  \caption{Fractional contribution of partonic channels to the total result at
    different perturbative orders, for the \WPC process, differential in
    $p_{T,j_c}$. The three columns correspond to different setups: OS$-$SS
    excl.\ (left), OS+SS excl.\ (middle), SS excl.\ (right). The three rows
    correspond to different perturbative orders: LO (top), NLO (middle), NNLO
    (bottom).}
  \label{fig:CHB_P_ptc}
\end{figure}

In order to investigate how the overall picture is affected by different
kinematical regions of phase space, we also investigate selected differential
distributions.
We focus on the \yl and \ptc observables, and we consider the fractional
contribution of each individual channel at each perturbative order for each bin
of the corresponding differential distributions.
The results are shown in Figure~\ref{fig:CHB_M_yl} (\yl in \WMC),
Figure~\ref{fig:CHB_P_yl} (\yl in \WPC), Figure~\ref{fig:CHB_M_ptc} (\ptc in
\WMC) and Figure~\ref{fig:CHB_P_ptc} (\ptc in \WPC).
In each figure, we plot the contribution of each partonic channel normalised to
the total at LO (1$^\text{st}$ row from the top), total NLO (2$^\text{nd}$ row
from the top) and total NNLO (3$^\text{rd}$ row from the top).
The left and the middle columns are in the OS$-$SS excl.\ and OS+SS
excl.\ setups, respectively, whereas the right column is in the SS setup.
We chose to plot OS$-$SS excl.\ and OS+SS excl.\, in order to have a
complementary information to the one provided in
Tables~\ref{tab:CHB_WMC}--\ref{tab:CHB_WPC}.
Instead, in the SS column, one can better appreciate the difference between the
several curves, given that the dominant $gs(\sb)$ component is absent.


We first focus on the OS$-$SS excl.\ and OS+SS excl.\ setups. In all plots, we
notice the dominance of the $gs(\sb)$ channel, as already observed for the
fiducial cross sections. However, it can be seen that for large values of \yl
and \ptc, the fractional contribution of $gs(\sb)$ decreases, with the other
channels starting to contribute more.
In particular, in the \yl distribution, we observe that $gs(\sb)$ is always very
close to 1 for most of the rapidity range, except for $\yl \gtrsim 2.0$ where it
decreases to 0.8. The overall picture is only mildly affected by the
perturbative order.
In contrast, in the \ptc distribution, we note a sharp decrease of the $gs(\sb)$
contribution as \ptc increases: while at LO $gs(\sb)$ is around 0.9 for low-\ptc
values down to 0.8 for high-\ptc values, at NLO and NNLO it goes down to
0.5--0.6 for $\ptc \sim 400$~GeV.
Other channels then give a non-negligible contribution at high transverse
momenta: the $gq(\qb)$ and $s(\sb)q(\qb)$ channels both in the OS$-$SS and OS+SS
setup; the $c(\cb)q(\qb)$ channels only in the OS+SS setup.
Indeed, by comparing left columns (OS$-$SS) with the middle columns (OS+SS), the
effect of the OS$-$SS subtraction is evident, with the curve associated to
$c(\cb)q(\qb)$ close to zero on the left.
As for the $gg$ channel, its contribution mildly depends on \yl, being negative
and constant in the whole rapidity range. Instead, it peaks at low-\ptc at NLO
and NNLO (where the total cross section is larger), with a negligible
contribution at large transverse momenta.

It is also interesting to note how the individual channels behave between \WMC
and \WPC. For instance, already at LO, the behaviour of the $gq(\qb)$ channel
both at large rapidities and at large transverse momenta is different, with a
larger contribution of $gq(\qb)$ in \WMC. These kinematic regions mainly receive
contributions from PDFs at large momentum fraction; hence, the plots confirm
that the origin of the difference between \WMC and \WPC to be related to the
valence component of the $d$ PDF, which is absent for the $\bar{d}$ PDF.
Equally noteworthy is the difference in size between the $c(\cb)q(\qb)$ and the
$q(\qb)q(\qb)$ channels in \WPC and \WMC at NLO and NNLO in the OS+SS
excl.\ setup.

We now consider the SS plots i.e.\ the column on the right in
Figs.~\ref{fig:CHB_M_yl}--\ref{fig:CHB_P_ptc}.
We see that both $c(\cb)q(\qb)$ and $q(\qb)q(\qb)$ are equally dominant for
small rapidity values, with $c(\cb)q(\qb)$ becoming larger and $q(\qb)q(\qb)$
becoming smaller at large rapidities, both at NLO and NNLO.
The situation is similar in the \ptc distribution, but starting from $\ptc
\gtrsim 300$~GeV the $c(\cb)q(\qb)$ channel constitutes the totality of the SS
cross section, with the $q(\qb)q(\qb)$ near to zero.
It is likely that in these events at large-\ptc the SS $c$-parton comes directly
from the PDFs: if it were radiatively generated, then other channels would also
contribute.

Having scrutinized in detail how contributions to the cross sections are
distributed among the various channels, we now return to consider the bottom
panels of Fig.~\ref{fig:DISTR_flav_ptj1}. Namely, understanding why the
behaviour of the considered setups is so different between \WMC and \WPC in the
\ptc distribution. This will give us the opportunity to further investigate the
correlation between PDFs and cross sections for \WMC and \WPC.

\begin{figure}[t]
  \centering  
  \includegraphics[width=0.94\textwidth]{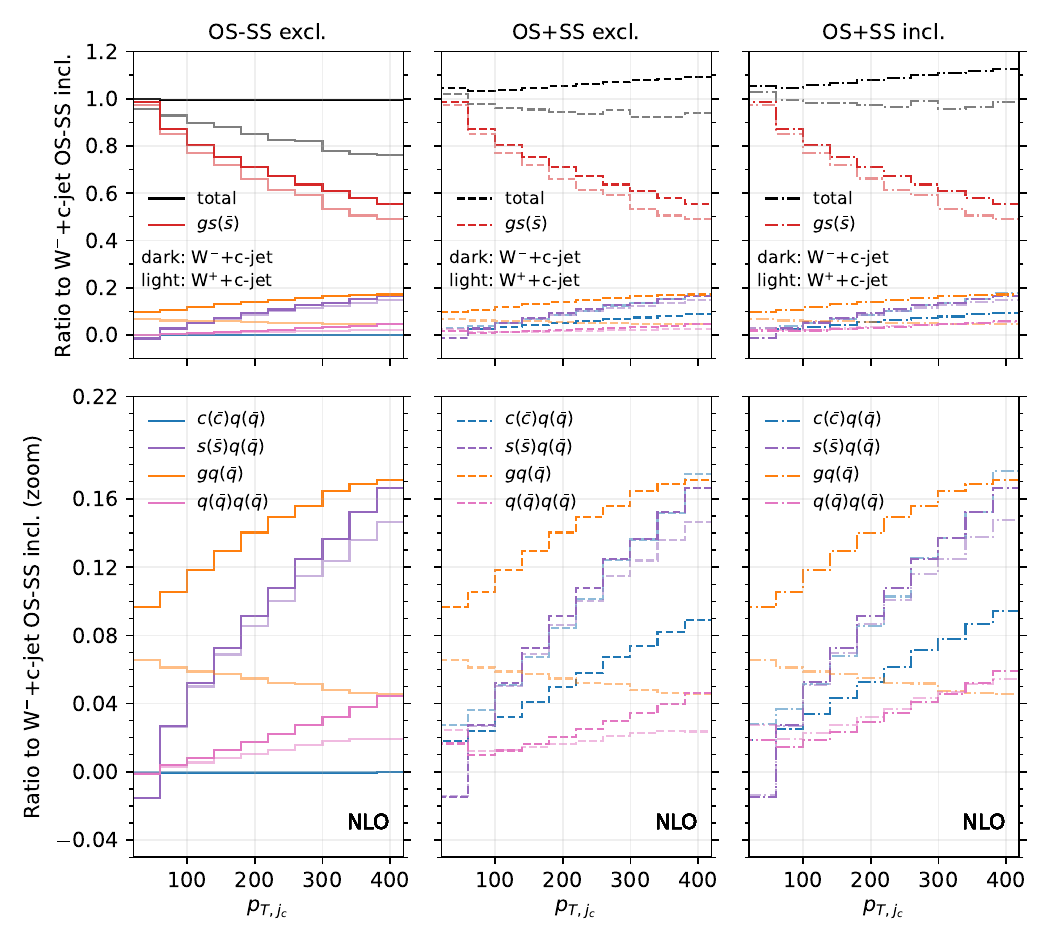}
  \caption{Analysis of the channels contributing to the \ptc distribution at NLO
    for OS$-$SS excl.(left), OS+SS excl.(middle) and OS+SS incl.(right).  All
    the curves are normalised to the \WMC OS$-$SS incl.\ NLO distribution. The
    lower panel is just a zoom of the upper panel. Darker colours refer to \WMC,
    lighter colours refer to \WPC.}
\label{fig:CHB_ALL_NLO}
\end{figure}

\begin{figure}[t]
  \centering  
  \includegraphics[width=0.94\textwidth]{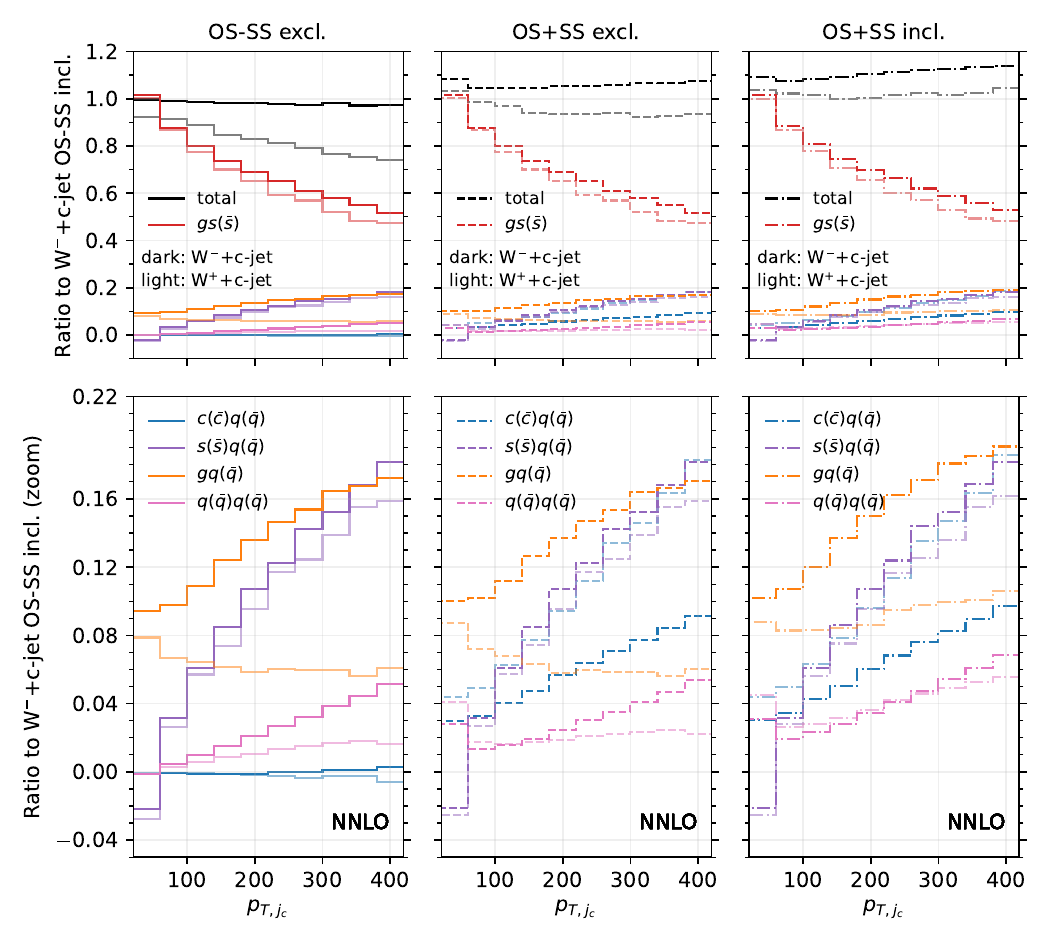}
  \caption{Analysis of the channels contributing to the \ptc distribution at
    NNLO for OS$-$SS excl.(left), OS+SS excl.(middle) and OS+SS incl.(right).
    All the curves are normalised to the \WMC OS$-$SS incl.\ NLO
    distribution. The lower panel is just a zoom of the upper panel. Darker
    colours refer to \WMC, lighter colours refer to \WPC.}
\label{fig:CHB_ALL_NNLO}
\end{figure}

Towards this aim, we consider again the \ptc distribution at NLO
(Figure~\ref{fig:CHB_ALL_NLO}) and at NNLO
(Figure~\ref{fig:CHB_ALL_NNLO}). However we now include curves with the
contributions of the most sizeable channels, and superimpose \WPC and \WMC on
the same plot, by choosing as common normalisation factor the \WMC OS$-$SS
incl.\ distribution.  In this way, we can determine the relative size of
contributions between \WPC and \WMC. The darker colours refer to \WMC, whereas
the lighter ones to \WPC.  We show results for the OS$-$SS excl.\ setup (left
frames), the OS+SS excl.\ setup (middle frames), the OS+SS incl.\ setup (right
frames).
The black curves in the upper left plots of Figures~\ref{fig:CHB_ALL_NLO}
and~\ref{fig:CHB_ALL_NNLO} coincide with the blue (NLO) and red (NNLO) curves in
the left plot in Fig.~\ref{fig:DISTR_flav_ptj1}.  Likewise, the gray curves in
the upper left plots of Figures~\ref{fig:CHB_ALL_NLO} and~\ref{fig:CHB_ALL_NNLO}
correspond to the blue and red curves in the right plot in
Fig.~\ref{fig:DISTR_flav_ptj1}, but they do not coincide as they have a
different normalisation.

We observe several important features. 
At NLO for both \WMC and \WPC, the difference between OS+SS excl.\ and OS$-$SS
excl.\ is driven by the $c(\cb)q(\qb)$ channel, and the difference between OS+SS
excl.\ and OS+SS incl.\ is driven by $q(\qb)q(\qb)$.
At NNLO, similar observations hold, with $gq(\qb)$ channel responsible for
further increasing the difference between OS+SS excl.\ and OS+SS incl.
Hence, explaining the lower panels of Fig.~\ref{fig:DISTR_flav_ptj1} amounts to
understanding why the $c(\cb)q(\qb)$, $q(\qb)q(\qb)$ and $gq(\qb)$ channels are
so different in size between \WMC and \WPC.

Starting from the $c(\cb)q(\qb)$ channel, from the discussion above we know that
in the high-\ptc region these events feature a SS $c$-parton coming directly
from the PDFs.
Therefore, the quark line coupling to the $W$-boson is unconstrained in terms of
flavour.
A typical diagram of such a configuration is displayed in Figure~\ref{fig:diags}
on the left.
The largest contribution in the large-$x$ region comes from the $d$ valence PDF
in the case of $W^{-}$ and from the $u$ valence PDF in the case of $W^{+}$. The
latter is approximately twice of the former, hence the factor of roughly 2
between the contribution of the $c(\cb)q(\qb)$ channel in OS+SS for \WMC and for
\WPC in the large-\ptc region is easily explained.

One can explain in a similar manner why the $q(\qb)q(\qb)$ and $gq(\qb)$
channels induce the difference between the incl.\ and the excl.\ setup, and why
such a difference is greater for \WPC compared to \WMC.
A typical diagram for $q(\qb)q(\qb)$ is shown in Fig.~\ref{fig:diags} on the
right. In this case, the charm is generated radiatively, hence we are summing
over the flavour combinations of the two incoming quarks.
Again the largest contributions in the large-$x$ region comes from the
$u\bar{d}$ channel in \WPC and from the $d\bar{u}$ channel in \WMC, so one
recover the factor of 2 in difference.
Similar considerations apply to the $gq(\qb)$ channels, which however features a
secondary pair of charm quarks only starting from NNLO.

\section{Conclusions}
\label{sec:concl}

In this paper, we presented a new calculation of $W$+charm-jet production up to
NNLO in QCD. We employed a new flavour-dressing procedure~\cite{Gauld:2022lem}
to define charm-jets in an IRC safe manner. Our results confirm an earlier
calculation~\cite{Czakon:2020coa,Czakon:2022khx}, applied to the kinematics of a
recent CMS measurement~\cite{CMS:2023aim}. A detailed decomposition into
different partonic channels demonstrated that the predominant contribution from
initial states containing strange quarks is maintained in most kinematical
distributions even when higher-order corrections are included.  The efficiency
of the OS$-$SS subtraction in removing contributions from secondary charm
production is clearly demonstrated by the channel decomposition. This
decomposition also explains the consistently larger magnitude of the \WMC over
\WPC cross sections to be due to contributions from CKM-suppressed $d$-valence
quark initiated processes.

Our results demonstrate the practical application of flavour
dressing~\cite{Gauld:2022lem} in NNLO QCD predictions. They will enable the
usage of $W$+charm-jet production observables in future global NNLO PDF fits and
thus enable a precise flavour composition of the quark content of the nucleon.

\acknowledgments
GS thanks the Institute for Theoretical Physics at ETH for hospitality during
the course of this work.  This research was supported in part by the UK Science
and Technology Facilities Council under contract ST/X000745/1, by the Swiss
National Science Foundation (SNF) under contracts 200021-197130 and
200020-204200 and by the European Research Council (ERC) under the European
Union's Horizon 2020 research and innovation programme grant agreement 101019620
(ERC Advanced Grant TOPUP).  Numerical simulations were facilitated by the High
Performance Computing group at ETH Z\"urich and the Swiss National
Supercomputing Centre (CSCS) under project ID ETH5f.

\bibliography{wcjet}

\end{document}